\documentclass[a4paper,11pt]{article}
\pdfoutput=1
\usepackage{jheppub} 
\usepackage{amsmath,amssymb,amsthm,amscd,graphicx}
\usepackage{psfrag}
\usepackage{subfig}
\usepackage{braket}
\input epsf.sty

\theoremstyle{definition}

\newcommand{\CC}{{\cal C}}

\newcommand{\CF}{{\cal F}}

\newcommand{\CN}{{\cal N}}

\def\IZ{{\mathbb Z}}

\def\IP{{\mathbb P}}

\def\l{\ell}
\newcommand{\tr}{\mathop{\rm Tr}\nolimits}
\newcommand{\re}{{\rm e}}
\newcommand{\ri}{{\rm i}}
\newcommand{\rd}{{\rm d}}

\newcommand{\Tr}{\mathop{\rm Tr}\nolimits}

\newcommand{\Vol}{\mathop{\rm Vol}\nolimits}
\newcommand{\pd}{\partial}
\def\cO{\mathcal{O}}
\def\({\left(}
\def\){\right)}

\newcommand{\be}{\begin{equation}}
\newcommand{\ee}{\end{equation}}
\newcommand{\ba}{\begin{aligned}}
\newcommand{\ea}{\end{aligned}}
\newcommand{\ben}{\begin{eqnarray}\displaystyle}
\newcommand{\een}{\end{eqnarray}}

\newcommand{\sectiono}[1]{\section{#1}\setcounter{equation}{0}}

\newdimen\tableauside\tableauside=1.0ex
\newdimen\tableaurule\tableaurule=0.4pt
\newdimen\tableaustep
\def\phantomhrule#1{\hbox{\vbox to0pt{\hrule height\tableaurule width#1\vss}}}
\def\phantomvrule#1{\vbox{\hbox to0pt{\vrule width\tableaurule height#1\hss}}}
\def\sqr{\vbox{%
  \phantomhrule\tableaustep
  \hbox{\phantomvrule\tableaustep\kern\tableaustep\phantomvrule\tableaustep}%
  \hbox{\vbox{\phantomhrule\tableauside}\kern-\tableaurule}}}
\def\squares#1{\hbox{\count0=#1\noindent\loop\sqr
  \advance\count0 by-1 \ifnum\count0>0\repeat}}
\def\tableau#1{\vcenter{\offinterlineskip
  \tableaustep=\tableauside\advance\tableaustep by-\tableaurule
  \kern\normallineskip\hbox
    {\kern\normallineskip\vbox
      {\gettableau#1 0 }%
     \kern\normallineskip\kern\tableaurule}%
  \kern\normallineskip\kern\tableaurule}}
\def\gettableau#1{\ifnum#1=0\let\next=\null\else
\squares{#1}\let\next=\gettableau\fi\next}

\def\l{\ell}

\newcommand{\figref}[1]{Fig.~\protect\ref{#1}}

\preprint{DESY 14-182}

\title{\boldmath Quantization conditions and functional equations in ABJ(M) theories}

\author{Alba Grassi$^a$,}
\author{Yasuyuki Hatsuda$^b$ }
\author{and Marcos Mari\~no$^a$}

\affiliation{
$^a$D\'epartement de Physique Th\'eorique et Section de Math\'ematiques,\\
Universit\'e de Gen\`eve, Gen\`eve, CH-1211 Switzerland\\
\\
$^b$DESY Theory Group, DESY Hamburg,\\ 
Notkestrasse 85, D-22603 Hamburg, Germany\\
\\}

\emailAdd{alba.grassi@unige.ch, yasuyuki.hatsuda@desy.de, marcos.marino@unige.ch}

\abstract{The partition function of ABJ(M) theories on the three-sphere can be regarded as the canonical partition function of an ideal 
Fermi gas with a non-trivial Hamiltonian. We propose an exact expression for the spectral determinant of this Hamiltonian, 
which generalizes recent results obtained in the maximally supersymmetric case. As a consequence, we find an exact
WKB quantization condition determining the spectrum which is in agreement with numerical results. In addition, 
we investigate the factorization properties and functional equations for our conjectured spectral determinants.
These functional equations relate the spectral determinants of ABJ theories with consecutive ranks of gauge groups
but the same Chern-Simons coupling.
}

\begin{document}

\maketitle
\flushbottom

\sectiono{Introduction}

In the last years, there has been a lot of progress in understanding ABJ(M) theory \cite{abjm, abj}.
In \cite{kwy} the partition function of ABJ(M) theory on the three-sphere was reduced to a matrix integral which turned out to be 
closely related to topological strings on local $\mathbb{P}^1\times \mathbb{P}^1$ \cite{mpabjm}.
In  \cite{dmp} the connection with topological strings was used to compute recursively the full 't Hooft $1/N$ expansion, which by the AdS/CFT correspondence corresponds to the 
genus expansion of a dual type IIA superstring theory. In order to understand the M-theory lifting of this result, one has to study ABJM theory in a different regime, usually 
called the M-theory regime or M-theory expansion, in which $N$ is large but the coupling constant is fixed. The study of the matrix models computing partition functions of Chern--Simons--matter theories 
in the M-theory regime was initiated in \cite{hkpt}, where the strict large $N$ limit was solved for a large class of theories. 

In \cite{mp}, a different approach was proposed to study the M-theory regime 
of ABJM theory and related models. In this approach, the partition function of ABJ(M) is interpreted as the partition function of an ideal Fermi gas. The M-theory limit corresponds to the thermodynamic limit of 
this gas, and the coupling constant of ABJM theory becomes Planck's constant. The Fermi gas formulation of ABJM theory has been intensively studied in 
\cite{mp,hmo,py, hmo2,cm,hmo3,hmmo}, 
leading to an exact expression for the the partition function of ABJ(M) theory which resums the 't Hooft expansion and includes as well non-perturbative, large $N$ instanton corrections \cite{dmpnp}. 

An important aspect of the Fermi gas approach is that, since we are dealing with an ideal gas, 
all the physics of the problem is encoded in the spectrum of the one-particle Hamiltonian. Therefore  one should be able to reproduce 
the results of \cite{hmmo} by studying the spectral problem associated to the Fermi gas of ABJ(M) theory. Conversely, the exact expression for the partition function should encode 
all the information about the spectrum of the Fermi gas. In \cite{km, kallen}, a WKB quantization condition for ABJ(M) theory has been proposed by studying the relation between the spectrum 
of the Hamiltonian and the thermodynamics of the Fermi gas. This condition turns out to be exact in the cases in which the theory has maximal supersymmetry 
\cite{cgm}, but it needs additional corrections in the general case, as it was recently pointed out in \cite{huang} (see also \cite{hw}) by a detailed numerical analysis.

One of the key results of \cite{cgm} is that, in the maximally supersymmetric cases, one can write an explicit expression for the grand canonical partition function 
of the Fermi gas, which is nothing but the spectral determinant of the Hamiltonian. This expression involves a Jacobi theta function, and the spectrum can be read from the vanishing locus of this 
theta function. In this paper we generalize the results of \cite{cgm} to ABJ(M) theories with $\CN=6$ supersymmetry. We write a general formula for the spectral determinant of 
these theories, which involves now a generalization of the theta function. In particular, we derive an exact WKB quantization condition for the spectrum. The quantization condition 
proposed in \cite{km, kallen} is only an approximation to the exact quantization condition, and in general it receives corrections that we can compute 
analytically in this paper. Our general result explains why the quantization conditions of \cite{km,kallen} are valid in the maximally supersymmetric cases. It reproduces the 
corrections found numerically in the case of ABJM theory in \cite{huang}, and we also test these corrections in detail against explicit calculations of the spectrum in both ABJM theory and ABJ theory. 

As it was emphasized in the companion paper \cite{ghm}, where we studied the implications of these ideas for topological string theory, the quantization condition 
is obtained as a corollary of a stronger result, namely a conjectural exact expression for the spectral determinant. This expression was tested in detail in \cite{cgm} in the 
maximally supersymmetric case, where it was shown that it reproduces the values for the partition functions calculated in \cite{hmo,py,abj-moriyama,honda-o}. In the general case with $\CN=6$ supersymmetry 
our conjecture for the spectral determinant is more difficult to verify, since this involves a resummation of the Gopakumar--Vafa expansion of the topological string free energy \cite{gv}. 
However, we are able to perform this resummation in one special case (ABJM theory with $k=4$), and we obtain a generating functional for the partition functions of this theory in full 
agreement with existing calculations \cite{hmo2}. 

In exactly solvable cases, spectral determinants enjoy very interesting properties. They can be factorized according to the parity of the eigenfunctions, 
and they satisfy functional equations (see for example 
\cite{voros,dt}). In this paper, we initiate the study of such properties in the spectral problem of ABJ(M) theory. We find for example an explicit factorization of the spectral determinant 
in the maximally supersymmetric case $k=1$, as well as conjectural functional equations akin to those found in \cite{voros, dt} in Quantum Mechanics. 
These functional equations relate the spectral determinants of ABJ theories with consecutive ranks of gauge groups.
In particular, if the Chern-Simons levels are odd, 
these equations determine all the ABJ spectral determinants
from the ABJM ones via the Seiberg-like duality of ABJ theory \cite{abj}.

This paper is organized as follows. In section 2 we review some general aspects of the Fermi gas formalism. In section 3 we introduce the spectral determinant and the generalized theta function 
associated to the ABJ(M) grand potential, and we deduce the exact quantization condition for the energy levels 
by looking at the zeros of this generalized theta function. We also give strong numerical evidence in support of our computations.
In section 4 we study an example with $\mathcal{N}= 6$ supersymmetry and we show that the full genus expansion can be resummed into an explicit function on the moduli space.
In section 5 we discuss the factorization of the spectral determinant according to the parity of the energy levels and in section 6 we give evidence for some functional identities.
In section 7 we draw some conclusions.
There are also three appendices.
In appendix \ref{annexj} and \ref{appendixsd} we give some details for computations appearing in section 4 and in section 5.  
In appendix \ref{appendixnum} we explain the numerical technique used to compute the spectrum.

\sectiono{The Fermi gas approach to ABJ(M) Theory}

The ABJ(M) theory \cite{abjm,abj} is an $\mathcal{N}=6$ superconformal Chern-Simons-matter theory with  gauge group $U(N_1)_k\times U(N_2)_{-k}$.
It consists in two Chern-Simons nodes, with  couplings $k$ and $-k$, respectively, together with four hypermultiplets in the bifundamental representation.
By using localization techniques it is possible to reduce the ABJ(M) partition function on $\mathbb{S}^3$ to the following matrix integral \cite{kwy}:
\be
\label{ABJZ}
\ba
&Z(N_1, N_2, k)\\
&={\ri^{-\frac{1}{2}(N_1^2-N_2^2)}\over N_1! N_2!} \int \prod_{i=1}^{N_1}{ \rd \mu_i  \over 2\pi} \prod_{j=1}^{N_2} {\rd \nu_j \over 2\pi}
 {\prod_{i<j} \left( 2 \sinh \left( {\mu_i -\mu_j \over 2}\right) \right)^2 \left(2 \sinh \left( {\nu_i -\nu_j \over 2}\right) \right)^2 \over 
\prod_{i,j}  \left(2 \cosh \left( {\mu_i -\nu_j \over 2}\right) \right)^2} \re^{-{\ri k \over 4 \pi}\left(  \sum_i \mu_i^2 -\sum_j \nu_j^2\right)}. 
\ea
\ee
When $N_1=N_2=N$ the above matrix integral can be also written as \cite{kwytwo,mp}
\be
\label{tanh-form}
Z (N,k)={1\over N!}  \int \prod_{i=1}^N {\rd x_i \over 4 \pi k}  {1\over 2 \cosh {x_i \over 2} } \prod_{i<j} \left( \tanh \left( {x_i - x_j \over 2 k } \right) \right)^2. 
\ee
These matrix integrals can be studied in the conventional 't Hooft expansion \cite{mpabjm, dmp}. In \cite{mp} it was pointed out that, to fully understand the non-perturbative effects, one has to go beyond the 't Hooft $1/N$ expansion and study the M-theory regime of (\ref{ABJZ}). In this regime, the ranks of the gauge groups are large but the coupling $k$ is fixed. To study 
this regime it is convenient to use the Fermi gas approach \cite{mp} in which we rewrite the matrix integral
as the canonical partition function of a one-dimensional ideal Fermi gas. In this approach, the Chern-Simons coupling $k$ plays the role of the Planck constant:
\be \hbar=2 \pi k. \ee  
The Fermi gas formulation of the ABJ matrix integral was proposed in \cite{ahs,honda} where (\ref{ABJZ}) was written as
\be 
\label{abj-fermi}
Z(N,N+M,k)=\re^{\ri \vartheta(N,M,k)}Z_{\rm CS}(M,k) \hat Z(N,k; M),
\ee
and we used the notation 
\be N=N_1, \quad M=N_2-N_1.
\ee
In the following we will suppose that
\be 
k\geq 0, \quad M\geq 0. 
\ee\
 The phase factor appearing in (\ref{abj-fermi}) is given by
\be
\re^{\ri \vartheta(N,M,k)}=\ri^{NM}\re^{-{\ri \pi \over 6 k}M (M^2-1)},\ee
and $Z_{\rm CS}(M,k)$ is the $U(M)$ Chern-Simon partition function on  $\mathbb{S}^3$ \cite{witten-cs}:
\be\label{ZCS} Z_{\rm CS}(M,k)
=   k^{-\frac{M}{2}}  \prod_{s=1}^{M-1} \left( 2\sin{\frac{\pi s}{k} } \right)^{M-s}.\ee
The factor $\hat Z(N,k; M)$ has the form, 
\be\label{abjZ}
\hat Z(N,k; M)= {1\over N!} \sum_{\sigma \in S_N}\int \rd^N x \prod_{i=1}^N\rho(x_i,x_{\sigma(i)}).  
\ee
The function $ \rho(x_1,x_2)$ is given as
\be \label{rhoabj}
\rho(x_1,x_2)= {1 \over 2 \pi k}{V_M ^{1/2}(x_1) V_M^{1/2}(x_2)\over 2 \cosh\left({x_1-x_2 \over 2k} \right)},\ee
where the function $V_M(x)$ is given by 
\be
V_M (x)={1\over \re^{x/2} + (-1)^M \re^{-x/2}} \prod_{s=-{M-1\over 2}}^{M-1\over 2} \tanh {x+ 2 \pi \ri s\over 2 k}. 
\label{Vabj}
\ee
One can verify that this function is real and positive. When $M=0$, the ABJ partition function becomes the partition function of ABJM theory given in (\ref{tanh-form}):
 \be
 Z(N, N, k)= \hat Z(N, k;0)= Z(N,k). 
 \ee
 %

%
%
%
%
%
 %

 The function (\ref{rhoabj}) can be interpreted as a canonical density matrix, 
\be
 \langle x_1| \hat \rho | x_2\rangle= \rho (x_1,x_2), \qquad \hat \rho =\re^{-\hat H}, 
\ee
where $\hat H$ is the one-particle Hamiltonian. In this picture, (\ref{abjZ}) is then the canonical partition function of an ideal 
Fermi gas of $N$ particles with Hamiltonian $\hat H$. 
Mathematically, the density matrix (\ref{rhoabj}) is given by a positive-definite, Hilbert-Schmidt integral kernel. The spectrum $E_n$ of the associated Hamiltonian $\hat H$ is 
then determined by the integral equation
\be \label{inteq} \int_{-\infty}^{\infty} \rho (x_1,x_2)\phi_n(x_2) \text{d} x_2= \re^{-E_n}\phi_n(x_1), \quad n \geq 0,\ee
where we have ordered the eigenvalues as
\be 
E_0  <  E_1 <  E_2 < \cdots 
 \ee
 %


%
\begin{figure}
\center
\includegraphics[height=6cm]{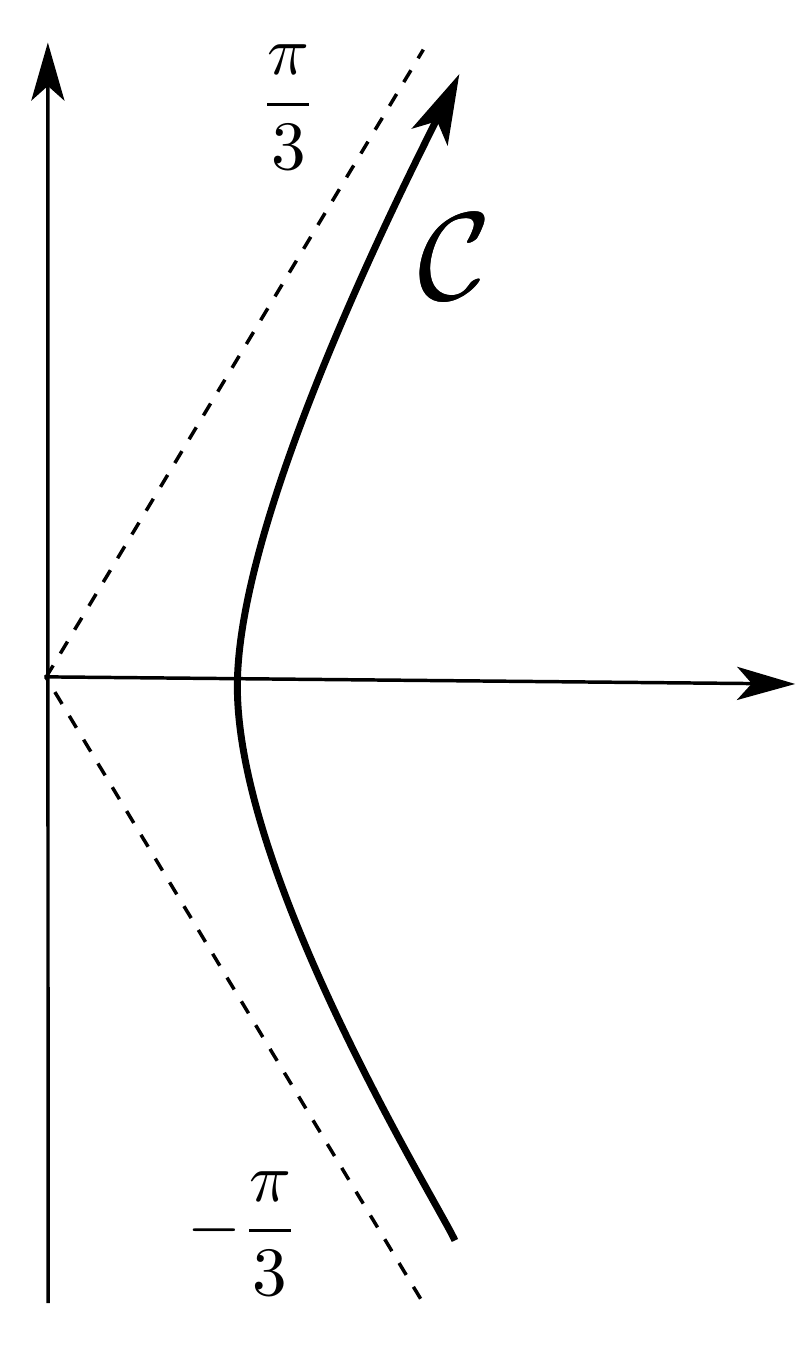}  
\caption{The standard Airy contour $\CC$  used to compute the canonical partition function from the modified grand potential.}
\label{airy-c}
\end{figure}

As it is well-known, ideal quantum gases are better studied in the grand canonical ensemble. The grand canonical partition function 
is defined by 

\be\label{spectdet}
\Xi(\kappa, k,M)=\det \left( 1+ \kappa \hat \rho \right)=\prod_{n=0}^\infty \left(1+ \kappa \re^{-E_n} \right),
\ee
where
 \be \kappa= \re^{\mu}\ee
is the fugacity. 
We will use  the notation $\Xi(\kappa, k, M)$ and $\Xi(\mu, k, M)$ interchangably.  When $M=0$ we will write
\be \Xi(\kappa, k)=\Xi(\kappa, k, 0).\ee 
The grand canonical partition function can be also interpreted as the {\it spectral determinant} (or Fredholm determinant) 
of the operator $\hat \rho$. Since this operator is positive-definite and Hilbert--Schmidt, it is of trace class and therefore its spectral determinant 
is well-defined\footnote{Note that our spectral determinant involves the spectrum of $\re^{-\hat H}$, rather than the spectrum of $\hat H$ itself, as in other definitions 
of spectral determinants \cite{voros,dt, cvitanovic}. Our definition is more convenient from the point of convergence properties, since it does not require regularization.}. 
It has two important properties that we will use later on. 
The first one is that $\Xi(\kappa, k, M)$ is an entire function of the the fugacity \cite{simon}, 
and the second one is that one can read off the physical energy spectrum by looking at the zeros of (\ref{spectdet}).
Indeed it is easy to see from the definition (\ref{spectdet}) that
\be 
\label{epi} 
\Xi({E+\ri \pi}, k,M)\ee
has simple zeros for 
\be E=E_n .\ee
The spectral determinant can be also regarded as a generating function for the partition functions $\hat{Z}(N,k, M)$:
\be 
\label{gen-fun}
\Xi(\kappa, k,M)=1+\sum_{N \ge 1}  \hat{Z}(N,k,M) \kappa^{N }.
\ee
The grand potential is defined by the usual formula, 
\be \label{JXi} \mathcal{J} (\mu, k,M)=\log \Xi(\kappa, k, M). 
\ee
Its power series expansion around $\kappa=0$ 
\be
\mathcal{J} (\mu, k,M)=\sum_{\ell \ge 1} {(-1)^{\ell-1} \over \ell} Z_\ell \kappa^\ell
\ee
involves the {\it spectral traces} of the canonical density matrix, 
\be
Z_\ell= \tr \, \hat \rho^\ell=\sum_{n\geq 0} \re^{-\ell E_n}. 
\ee

In the context of ABJ(M) theory it is convenient to use the \textit{modified} grand potential $J(\mu,k, M)$, 
which was introduced in \cite{hmo2}. It is related to the partition function by 
\be \label{ZJ} \hat Z(N,k,M)=\int_{\mathcal{C}}{\rm{d} \mu\over 2 \pi \ri }\re^{J(\mu,k,M)- N \mu}, 
\ee
where $\mathcal{C}$ is the standard Airy contour shown in Figure \ref{airy-c}. 
The standard  and the modified grand potentials are related via
\be\label{JJ} \re^{\mathcal{J} (\mu, k,M)}=\sum_{n \in \mathbb{Z}} \re^{J(\mu+2 \pi \ri n,k,M)}.\ee
The modified grand potential of ABJM theory was determined in a series of works \cite{mp, hmo, hmo2, cm, hmo3, hmmo}, and it 
can be written down in terms of the standard and refined topological strings on local $\mathbb{P}^1 \times \mathbb{P}^1$. This result was extended to 
ABJ theory in \cite{honda-o,abj-moriyama}. One has the following result: 
\be
\label{gpmueff}
J(\mu, k,M)=J^{(\rm p)}(\mu_{\rm eff},k,M)+J^{\rm WS} (\mu_{\rm eff},k, M)+ \mu_{\rm eff}
\widetilde{J}_b(\mu_{\rm eff},k,M)+\widetilde{J}_c(\mu_{\rm eff},k,M).
\ee
The perturbative piece $J^{(\rm p)}$ is a cubic polynomial in $\mu$: 
\be
J^{(\rm p)}(\mu,k,M)= {C(k) \over 3} \mu^3 + B(k,M) \mu + A(k,M), 
\ee
where 
\be
\label{CB}\ba 
 C(k)= {2\over \pi^2 k}, \qquad
B(k,M)={1\over 3k}-{k\over 12}+{k \over 2}\left( {M \over k}-{1\over 2}\right)^2.\ea
\ee
The constant term is given by
\be \label{Akm}\ba
A(k,M)= -\log{|Z_{\rm CS} (M,k)  |}
         +\frac{2\zeta(3)}{\pi^2 k}\left(1-\frac{k^3}{16}\right)+\frac{k^2}{\pi^2} \int_0^\infty \rd x \frac{x}{\re^{k x}-1}\log(1-\re^{-2x})\\
  \ea  \ee
  where $Z_{\rm CS} (M,k) $ is the same as in (\ref{ZCS}).  
In particular for $M=0$ we have 
\be 
Z_{\rm CS}(0,k)=1,
\ee
and we recover the constant map contribution of ABJM theory \cite{hanada,hatsuda-o}.
The exact values of this constant map contribution for arbitrary integral $k$ are found in \cite{hatsuda-o}.
The effective chemical potential $\mu_{\rm eff}$ was introduced in \cite{hmo3} to take into account 
bound states of worldsheet instantons and membrane instantons. It is given by 
\be
\label{mueff-mu}
\mu_{\rm eff}= \mu - \frac{1}{2} \sum_{\ell=1}^\infty  (-1)^{M \ell} \hat a_\ell(k) \re^{-2\ell \mu}. 
\ee
In \cite{hmmo}, it was shown that the coefficients $\hat a_\ell(k)$ are determined by the coefficients of the so-called quantum mirror map of local $\IP^1 \times \IP^1$, introduced in \cite{acdkv}. 
For the first few terms we have
\be \ba
\hat a_1(k)&=2(q^{1/2}+q^{-1/2}) ,\\
\hat a_2(k)&=5(q+q^{-1})+8 ,\\
\hat a_3(k)&=2(q^{5/2}+q^{-5/2})+\frac{62}{3}(q^{3/2}+q^{-3/2})+44(q^{1/2}+q^{-1/2}) ,
\ea \ee
and we denoted
\be q=\re^{\ri  \pi k}.\ee
When $k$ is an integer, the effective chemical potential can be written in closed form 
\cite{honda-o}
\be 
\mu_{\rm eff} = \left\{ \begin{matrix}\mu  -2(-1)^{\frac{k}{2}-M}\re^{-2\mu} \ _4 F_3 \left( 1,1,\frac{3}{2},\frac{3}{2},2,2,2; (-1)^{\frac{k}{2}-M}16\, \re^{-2\mu} \right) , \quad \text{if $k$ is even} \cr
\cr 
\mu +\re^{-4\mu} \ _4 F_3 \left( 1,1,\frac{3}{2},\frac{3}{2},2,2,2; -16\,  \re^{-4\mu} \right) , \hspace{3.6 cm} \text{if $k$ is odd}.\cr
\end{matrix} \right. 
\ee
The membrane part of the grand potential consists of two functions $\widetilde{J}_b(\mu_{\rm eff},k,M)$ and
$\widetilde{J}_c(\mu_{\rm eff},k,M)$, whose large $\mu$ expansion reads:
\begin{align}
 \label{Jme}
 \widetilde{J}_b(\mu_{\rm eff},k,M)=\sum_{\l=1}^\infty\widetilde{b}_\l(k)(-1)^{M \ell}\re^{-2\l\mu_{\rm eff}}, \qquad \widetilde{J}_c(\mu_{\rm eff},k,M)
=\sum_{\l=1}^\infty\widetilde{c}_\l(k)(-1)^{M \ell}\re^{-2\l\mu_{\rm eff}}. 
\end{align}
The coefficients $\widetilde b_\ell(k)$ are related to the quantum B-period of local $\mathbb{P}^1 \times \mathbb{P}^1 $ \cite{hmmo}, 
and can be expressed in terms of the refined BPS invariants
$N^{d_1,d_2}_{j_L,j_R}$ of this CY \cite{gv,ikv,ckk}, as
\be
\label{blj}
\widetilde{b}_\ell(k)=-\frac{\ell}{2\pi}\sum_{j_L,j_R}\sum_{\ell=dw}\sum_{d_1+d_2=d}N^{d_1,d_2}_{j_L,j_R}q^{\frac{w}{2}(d_1-d_2)}
\frac{\sin\frac{\pi kw}{2}(2j_L+1)\sin\frac{\pi kw}{2}(2j_R+1)}{w^2\sin^3\frac{\pi kw}{2}}. 
\ee
The particular combination of invariants appearing here involves only the so-called Nekrasov--Shatashvili limit \cite{ns} of the topological string free energy. 
The coefficients $\widetilde c_\ell(k)$ can be computed from the $\widetilde b_\ell(k)$ by using the relation conjectured in \cite{hmo3}
\begin{align}
\widetilde{c}_\l(k)=- k^2 \frac{\partial}{\partial k} 
\left(\frac{\widetilde{b}_\l(k)}{2\l k}\right). 
\label{bcrel}
\end{align}
The worldsheet part of the grand potential $J^{\rm WS} (\mu,k)$ takes the following form
\be \label{JWS}
J^{\rm WS} (\mu,k,M)= \sum_{m \ge 1} (-1)^m d_m (k,M) \re^{-4 m \mu/k},
\ee
where the coefficient $ d_m (k,M)$ can be expressed in terms of the Gopakumar--Vafa invariants 
$n_g^{d_1, d_2}$ of local $\IP^1 \times \IP^1$ \cite{gv}. It reads (see  also \cite{kallen})\footnote{The coefficients $d_m (k,M)$ differ from those in \cite{kallen} by a factor $(-1)^m$. }
\be 
d_m (k,M)=\sum_{g \geq 0}\sum_{dn=m}\sum_{d_1+d_2=d}
n_g^{d_1,d_2}\beta^{{d_2-d_1\over d} m}\left(2 \sin{2\pi n \over k}\right)^{2g-2}{1\over n},\ee
where \be \beta=\re^{-2\pi \ri M/k}.
\ee
These coefficients can also be expressed in terms of BPS invariants of local $\mathbb{P}^1 \times \mathbb{P}^1$
\be
d_m(k,M)=\sum_{j_L,j_R}\sum_{dn=m}\sum_{d_1+d_2=d}
 \frac{2j_R+1}{\sin^2 \frac{2\pi n}{k}}
\frac{\sin( \frac{4\pi n}{k}(2j_L+1) )}{n \sin \frac{4\pi n}{k}}N_{j_L,j_R}^{d_1,d_2}\beta^{{d_2-d_1\over d} m}.
\ee
Notice that both the worldsheet instanton contributions (\ref{JWS}) and the membrane instanton contributions (\ref{Jme}) have poles at rational value of $k$.
 However, as noted in \cite{honda-o}, the HMO cancellation mechanism of ABJM theory \cite{hmo2,hmmo} extends to ABJ theory, and 
 these poles cancel in the total sum. As a result, the modified grand potential is a well defined and finite quantity for any value of $k$.

\sectiono{Spectral determinant and quantization conditions}

The physical information on the ABJ(M) Fermi gas can be encoded in many different ways: in the spectrum of the Hamiltonian, in the 
spectral determinant, and in the modified grand potential. It is clear that these three objects are equivalent, but their relationship is not trivial. 
The purpose of this paper is to use the explicit answer for the modified grand potential of ABJ(M) theory in order to find a useful expression for the 
spectral determinant and to solve the spectral problem of the Hamiltonian. 
\begin{figure}
\center
{\includegraphics[height= 5 cm]{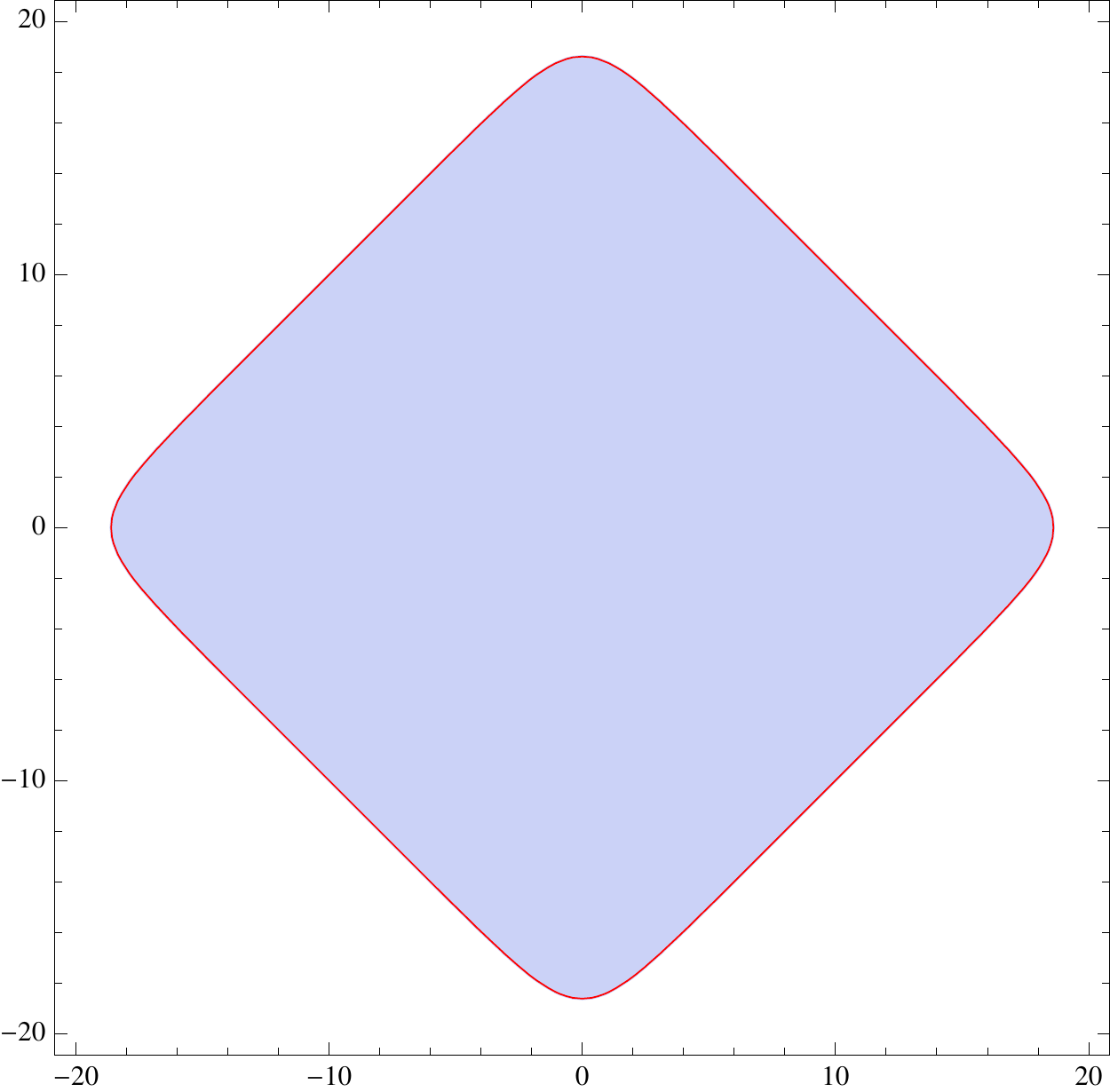}  }
\caption{ The region of phase space (\ref{pscl}) for ABJM theory, at large energy. }
\label{clvol}
\end{figure}

A natural starting point to find the spectrum is to use the Bohr--Sommerfeld approximation, which in this generalized 
setting goes as follows. Let ${\rm{Vol}}_{\rm cl}(E,M)$ be the classical phase space volume, i.e. 
the volume of the region
\be 
\label{pscl}
 \mathcal{R}_{\rm cl}(E)=\left\{\left( x,p\right) \in \mathbb{R}^2 :   \rho_{\rm cl} (x,  p) \leq \re^{-E} \right\}. \ee
Here we denoted by $\rho_{\rm cl} (x,  p)$ the classical limit\footnote{%
As mentioned in \cite{ahs}, the total ABJ partition function \eqref{abj-fermi} vanishes for $k<M$. This is because the Chern-Simons factor
$Z_\text{CS}(M,k)$ becomes zero in this regime. However the normalized partition function $\hat Z(N,k; M)$ defined by \eqref{abjZ}
is still non-zero even for $k<M$. Therefore one can consider the classical limit once going to 
the Fermi gas system. 
}
of the quantum operator $\hat  \rho$, which is given by 
\be  \rho_{\rm cl} (x,  p)=\re^{-T(p)-U(x,M)},
\ee
where\footnote{ Strictly speaking   $\rho_{\rm cl} (x,  p)$ is not fully classical because $U(x,M)$ still depends on the Planck constant when $M>0$.}
\be \ba T( p)= \log \left(2 \cosh {p \over 2} \right), \quad
U( x,M)=-\log \left( V_M(x) \right).
\ea \ee
The Bohr--Sommerfeld quantization condition reads
\be  
\label{bs}
\text{Vol}_{\rm cl}(E,k, M)=2 \pi \hbar \left(n +{1\over 2} \right), \qquad n=0, 1,2, \cdots
 \ee
For large values of the energy one finds
\be 
 \text{Vol}_{\rm cl}(E,k, M)\approx 8 E^2, 
 \ee
as shown for instance in Figure \ref{clvol}. 
Notice that the region $\mathcal{R}_{\rm cl}(E)$  has a finite volume, corresponding to the fact that 
the operator $\hat \rho$ has a discrete spectrum.
In \cite{mp} it was pointed out that the classical volume receives two types of quantum corrections, 
perturbative and non perturbative in $\hbar$, and 
there should be a fully ``quantum" version of the classical function $\text{Vol}_{\rm cl}(E,k, M)$ incorporating these corrections, which we will denote 
by $\text{Vol}(E, k, M)$. It is convenient to write the quantum volume as in \cite{km}, 
\be \text{Vol}(E, k, M)=\text{Vol}_{\rm{p}}(E,k, M)+\text{Vol}_{{\rm{np}}}(E,k, M), 
\ee
where $\text{Vol}_{\rm{p}}(E,k, M)$ contains the full series of perturbative corrections in $\hbar$, while $\text{Vol}_{{\rm{np}}}(E,k, M)$ 
contains  the non-perturbative corrections in $\hbar$. The Bohr--Sommerfeld quantization condition should be promoted to an {\it exact} 
WKB quantization condition of the form 
\be 
\label{bs-exact} \text{Vol}(E,k, M)=2 \pi \hbar \left(n +{1\over 2} \right), \qquad n=0, 1,2, \cdots
 \ee
 similar to what happens in some problems in ordinary Quantum Mechanics \cite{zj}. 
 
 Our goal is to extract the exact quantum volume from our knowledge of the 
 exact grand potential. A first attempt to do this was presented in \cite{km}. Although the strategy of \cite{km} leads to the correct 
 result in the case of $k=1,2$, it involves many technical difficulties in the study of the non-perturbative sector. 
Here we overcome these difficulties by using the approach of \cite{cgm}, where the quantum volume and the spectrum are computed by studying the zeros of the spectral determinant \footnote{As noted 
in \cite{cvitanovic}, p. 672, ``the smart way to determine the eigenvalues of linear operators is by determining zeros of their spectral determinants."}. 
Therefore, we will first find a convenient expression for the spectral determinant of these theories. 

In the case of maximally supersymmetric theories, it was shown in \cite{cgm} that the sum appearing in (\ref{JJ}) can be written in terms 
of Jacobi theta functions. It is easy to see, by a computation similar to the one presented in \cite{ghm}, that the spectral determinant for ABJ(M) theory 
(\ref{JJ}) is given by 
 \be 
 \label{spec-det}
 \Xi (\mu, k,M)=\re^{J(\mu,k,M)} \Theta (\mu, k ,M),
 \ee
where 
\be
\label{ggtf}
\ba
\Theta (\mu, k, M)= \sum_{n \in \IZ} \exp \biggl\{ & - 4\pi^2 n^2 C(k) \mu_{\rm eff} + 2 \pi \ri n \left( C(k) \mu_{\rm eff}^2+ B(k,M)
+\widetilde{J}_b (\mu_{\rm eff}, k,M)  \right)\\ 
& +  J_{\rm WS}(\mu_{\rm eff} + 2 \pi \ri n, k,M) - J_{\rm WS}(\mu_{\rm eff}, k,M)- {8 \pi^3 \ri n^3 \over 3} C(k) 
\biggr\}. 
\ea
\ee
We will call this function a {\it generalized theta function}.

As we noted in (\ref{epi}), the spectrum of energies in (\ref{inteq}) can be obtained by looking at the zeros of the spectral determinant, by setting 
 \be \mu=E + \pi \ri. 
 \ee
As it was found in \cite{cgm,ghm}, this involves looking at the zeros of the (generalized) theta function, and leads to a quantization condition of the form (\ref{bs-exact}) which incorporates all the perturbative and non-perturbative corrections to the Bohr--Sommerfeld condition (\ref{bs}). 
It is easy to see that
\be
\label{theta12}
\ba
\Theta(E+\pi \ri, k, M)= \re^{\zeta} \sum_{n \in \IZ} \exp \biggl\{  &- 4\pi^2 (n+1/2)^2 C(k) E_{\rm eff} 
 - {8 \pi^3 \ri (n+1/2)^3 \over 3} C(k) \\
 &+ 2 \pi \ri (n +1/2)\left( C(k) E_{\rm eff}^2+ B(k,M) +\tilde J_b (E_{\rm eff}+ \pi \ri, k,M) \right)\\
&+ f_{\rm WS}(E_{\rm eff}+ \pi \ri, n)-{1\over 2} f_{\rm WS} (E_{\rm eff}+ \pi \ri,-1)\ \biggr\}. 
\ea
\ee
In this equation we have introduced, in analogy with (\ref{mueff-mu}), the ``effective" energy
 \be
\label{Eeff-E}
E_{\rm eff}= E - \frac{1}{2} \sum_{\ell=1}^\infty  (-1)^{M \ell} \hat a_\ell(k) \re^{-2\ell E}. 
\ee
In addition, we have 
 \be
 \ba
 f_{\rm WS}(\mu, n) &=\sum_{m \ge 1} d_m (k,M)\left( \re^{-8 \pi \ri m  n /k}-1\right) (-1)^{ m} \re^{-4 m  \mu/k},
 \ea
 \ee
so that
 \be
f_{\rm WS} (E_{\rm eff}+ \pi \ri,-1)= 2 \ri  \sum_{m \ge 1} d_m (k,M)\sin {4 \pi m\over k}  (-1)^{ m} \re^{-4m E_{\rm eff} /k}. 
\ee
 We also have that 
 \be
\ba
\zeta&={2\over k} E_{\rm eff} -\pi \ri \left( {2\over \pi^2 k} E_{\rm eff}^2+ B(k,M) +\tilde J_b (E_{\rm eff}+ \pi \ri, k,M)\right) \\
&+{1\over 2} f_{\rm WS}(E_{\rm eff}+ \pi \ri,-1) + {2 \pi \ri  \over 3 k}. 
\ea
\ee
Note that, if we just take into account in the generalized theta function (\ref{theta12}) the terms with $n=0,-1$, we obtain the quantization condition
\be
\label{km-qc}
\cos \left(  \pi \Omega (E,k,M)   \right)=0, 
\ee
where
\be
\ba
\Omega (E,k,M) &=  
C(k) E_{\rm eff}^2+ B(k,M)- {\pi^2 C(k) \over 3} + \sum_{\l=1}^\infty\widetilde{b}_\l(k)(-1)^{M \ell}\re^{-2\ell E_{\rm eff}} \\
&\quad- {1\over  \pi }  \sum_{m \ge 1} (-1)^m d_m (k,M)  \sin \left( {4 \pi m \over k} \right) \re^{-4 m E_{\rm eff} /k}.
\ea
\ee
This is precisely the quantization condition proposed in \cite{km, kallen}. 
However, there will be in general corrections to this condition, due to the remaining terms in (\ref{theta12}). 
In order to take them into account systematically, let us call this correction $\lambda(E)$  and write the exact quantization condition as
\be \Omega (E, k,M)+\lambda(E)= s+{1\over 2}, \quad s=0,1,2, \dots. \label{exact-qc}
\ee
A straightforward calculation shows that $\lambda(E)$, which is non-perturbative in $k$ (i.e. in $\hbar$), is determined by the following equation 
\cite{ghm}
\be \label{Volc}
\sum_{n \geq 0} \re^{-\frac{8n(n+1)}{k}E_{\rm eff}}(-1)^n \re^{f_\text{c}(n)}\sin \left(
\frac{8\pi n(n+1)(2n+1)}{3k}+f_\text{s}(n)+2\pi \( n+\frac{1}{2}\)\lambda(E) \right)=0,
\ee
where
\be
\ba
f_\text{c}(n)&=\sum_{m \geq 1}(-1)^m d_m(k,M)\biggl[
\cos \( \frac{4\pi m(2n+1)}{k} \)-\cos \( \frac{4\pi m}{k} \) \biggr]\re^{-\frac{4m}{k}E_{\rm eff}},\\
f_\text{s}(n)&=\sum_{m \geq 1}(-1)^m d_m(k,M)\biggl[
\sin \( \frac{4\pi m(2n+1)}{k} \)-(2n+1)\sin \( \frac{4\pi m}{k} \) \biggr]\re^{-\frac{4m}{k}E_{\rm eff}}.
\ea
\ee
We also conclude from this analysis that the exact quantum volume is given by 
\be
\label{total-vol}
{\rm Vol}(E, k, M)= 2 \pi \hbar \left( \Omega (E, k,M)+\lambda(E) \right). 
\ee
Note that the perturbative part of this quantum volume is given precisely by the all-orders WKB perturbative contribution, encoded in the quantum B-period, 
\be
{1\over 2 \pi \hbar} {\rm Vol}_{\rm p}(E, k, M)=  C(k) E_{\rm eff}^2+ B(k,M)- {\pi^2 C(k) \over 3} + \sum_{\l=1}^\infty\widetilde{b}_\l(k)(-1)^{M \ell}\re^{-2\ell E_{\rm eff}}, 
\ee
while the non-perturbative contribution is given by 
\be
{1\over 2 \pi \hbar} {\rm Vol}_{\rm np}(E, k, M)=- {1\over  \pi }  \sum_{m \ge 1} (-1)^m d_m (k,M)  \sin \left( {4 \pi m \over k} \right) \re^{-4 m E_{\rm eff} /k} 
+ \lambda(E). 
\ee
The perturbative and the non-perturbative part are separately divergent when $k$ is rational, as noted in \cite{km}, but the total quantum volume 
(\ref{total-vol}) is smooth, since the singularities cancel as a consequence of the HMO mechanism (indeed, the quantum volume is obtained from 
the modified grand potential, which is singularity-free). The non-perturbative part is then needed to cancel the singularities in the WKB perturbative 
expansion, and it contains crucial information on the spectrum. For example, as shown in appendix \ref{annexj}, when $k$ is an integer, 
the finite part of $\tilde J_b (E_{\rm eff}+ \pi \ri, k,M)$ vanishes and the quantum volume is largely determined by the 
worldsheet instanton contribution. 

The correction $\lambda(E)$ can be computed analytically, in a series expansion in $\re^{-4E_{\rm eff}/k}$.
It is easy to see that $\lambda(E)$ is of the form 
\be
\label{le-series}
\lambda(E)=\sum_{\ell \geq 1} \lambda_\ell \, \re^{-\frac{4\ell+12}{k}E_{\rm eff}},
\ee
where the first few terms are explicitly given by
\be
\ba
\lambda_1&=\frac{1}{\pi} \sin (16x) ,\\
\lambda_2&=\frac{4}{\pi}\sin^2( 4x ) \sin ( 20x ) d_1,\\
\lambda_3&=\frac{8}{\pi}\sin^2( 4x ) \sin ( 24x )[ \sin^2 ( 4x )d_1^2-2\cos^2 ( 4x ) d_2 ],\\
\lambda_4&=\frac{4}{3\pi}\sin^2 (4x) \sin(28x) [ 3(d_1^3-2d_1d_2+3d_3)
-4\cos(8x) (d_1^3-3d_3)\\
&\quad+\cos(16x)(d_1^3+6d_1d_2+6d_3) ]
\ea
\ee
with 
%
\be
\ba
d_1&=d_1(k,M)=-\cos \( 2M x \) \csc^2 \( 2x \), \\
d_2&=d_2(k,M)=-\csc^2\( 2x \) -\frac{1}{2}\cos \( 4Mx \)\csc^2 \( 4x \),\\
d_3&=d_3(k,M)=-3\cos\( 2Mx \) \csc^2 \( 2x \)
-\frac{1}{3}\cos\( 6Mx \) \csc^2 \( 6x \), 
\ea
\ee
and we have denoted
\be x={\pi \over k}.
\ee
Note that, when $k=1,2,4$, we have that $f_{\rm s}(n)=0$, and the first term in the argument of the sine in (\ref{Volc}) is always a multiple of $\pi$. 
Therefore, the solution to (\ref{Volc}) is $\lambda(E)=0$, i.e. the correction vanishes and the quantization condition of \cite{km, kallen} is exact.

\begin{table}[t] 
\centering
\begin{tabular}{l l l}
&Energy levels for $k=3, M=0$&\\
\hline
Order &	$E_0$ & $E_1$ \\
\hline
$ \re^{-4 E /3}$ &    ${\underline{2.65}297702084083921}$ & \underline{4.68940}459079460092512108986442\\
$ \re^{-12 E /3}$ &  ${\underline{2.65156}164019289190}$ & \underline{4.6894013445}0544960666103687122\\
$\re^{-24 E /3}$ &   ${\underline{2.65156833}993530136}$ & \underline{4.6894013445703167}8330042507336\\
$\re^{-32 E /3}$ &   ${\underline{2.65156833716}940289}$ & \underline{4.6894013445703167756175}7482976\\
$\re^{-40 E /3}$ &   ${\underline{2.651568337168}75544}$ & \underline{4.68940134457031677561753154}101\\ 
$\re^{-52 E /3}$ &   ${\underline{2.651568337168857}61}$ & \underline{4.68940134457031677561753154681}\\
\hline
Numerical value &                        2.65156833716885755 &   4.68940134457031677561753154681
\\
\end{tabular}
\caption{The first two energy levels for $k=3$ and  $M=0$ calculated analytically from (\ref{exact-qc}). In the last line numerical values
evaluated by the method in appendix~\ref{appendixnum} are given. 
At each order of the approximation, we underline the digits which agree with the numerical result.  }
 \label{k03table}
\end{table}
\begin{table}[t] 
\centering
\begin{tabular}{l l l}
\\
&Energy levels for $k=5, M=0$&\\
\hline
Order &	$E_0$ & $E_1$ \\
\hline
$ \re^{-4 E /5}$ &    ${\underline{3.0}475013693}$ & \underline{5.793}53763401508120749977\\
$\re^{-16 E /5}$ &   ${\underline{3.0724}584475}$ & \underline{5.793694691}26135544218070\\
$\re^{-32 E /5}$ &   ${\underline{3.072435}9155}$ & \underline{5.793694691073381}73784939\\
$\re^{-48 E /5}$ &   ${\underline{3.07243583}57}$ & \underline{5.793694691073381584125}49\\
\hline
Numerical value &                         3.0724358360    &                      5.79369469107338158412559
\\
\end{tabular}
\begin{tabular}{l l l}
\\
&Energy levels for $k=6, M=0$&\\
\hline
Order &	$E_0$ & $E_1$ \\
\hline
$ \re^{-4 E /6}$ &    ${\underline{3.2}1322311}$ & \underline{6.231}11654150891713732\\
$\re^{-16 E /6}$ &   ${\underline{3.23}510192}$ & \underline{6.2314199}9560231213049\\
$\re^{-32 E /6}$ &   ${\underline{3.23464}705}$ & \underline{6.2314199801895}4896785\\
$\re^{-48 E /6}$ &   ${\underline{3.234644}06}$ & \underline{6.231419980189532863}12\\
\hline
Numerical value &                         3.23464413    &                      6.23141998018953286330       
\end{tabular}
\caption{The first two energy levels for $k=5,6$ and  $M=0$ calculated analytically from (\ref{exact-qc}). 
}
 \label{k0table}
\end{table}%
 \begin{figure} \begin{center}
 {\includegraphics[scale=0.6]{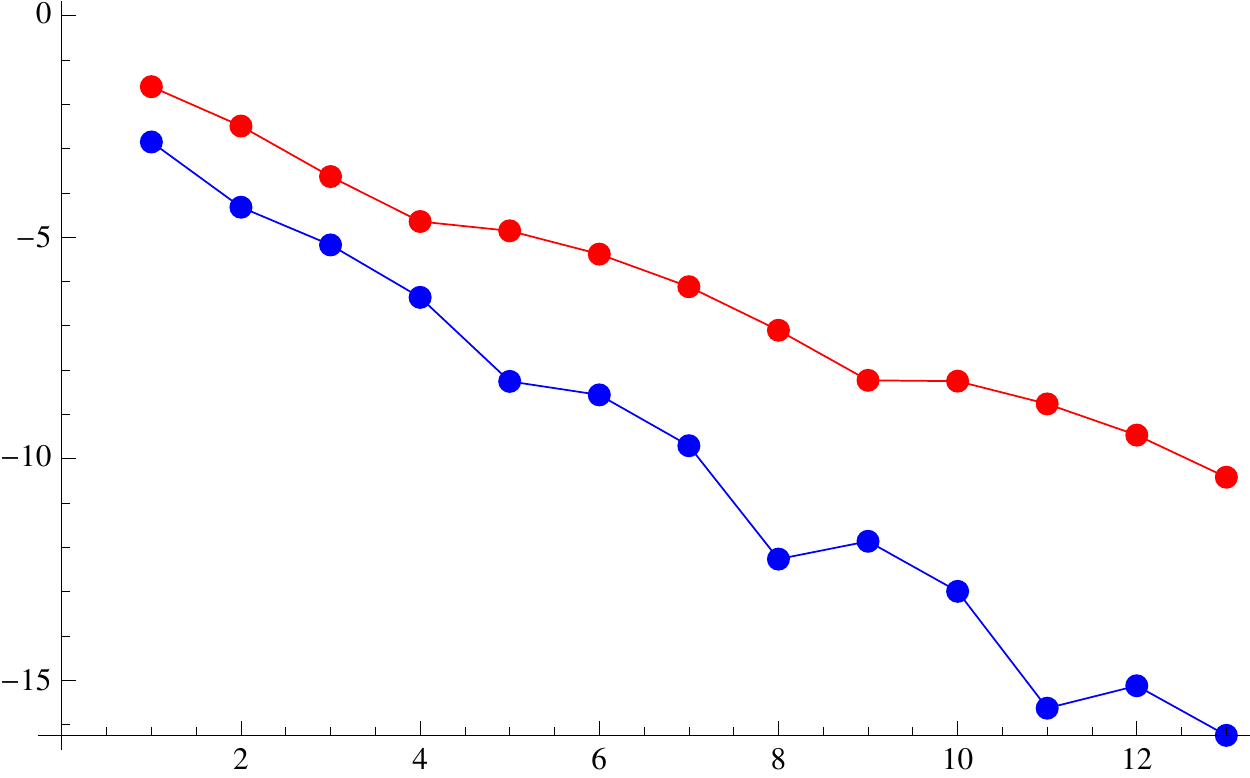} }
\caption{The difference $\Delta_0^{(k)}(m)$ (\ref{diff}) as a function of $m$, the number of instantons included in the computation. The line on the bottom (in blue) gives $\Delta_0^{(3)}(m)$, 
while the line on the top (in red) gives $\Delta_0^{(5)}(m)$.}
 \label{Figos}
  \end{center}
\end{figure}
\begin{table}[t] 
\centering
\begin{tabular}{l l l}
&Energy levels for $k=3, M=1$&\\
\hline
Order &	$E_0$ & $E_1$ \\
\hline
$ \re^{-8 E /3}$ &   ${\underline{3.48669}71392036144}$ & \underline{5.19022910}2060787166657584\\
$ \re^{-12 E /3}$ & ${\underline{3.4866953}311076197}$ & \underline{5.190229100088}979204304574\\
$\re^{-20 E /3}$ &  ${\underline{3.4866953293248}867}$ & \underline{5.1902291000888347}84550474 \\
$\re^{-28 E /3}$ & ${\underline{3.48669532933487}93}$  & \underline{5.1902291000888347966169}34\\
\hline
Numerical value & 3.4866953293348782  & 5.190229100088834796616924
\end{tabular}
\caption{The first two energy levels for $k=3, M=1$ calculated analytically from (\ref{exact-qc}). 
}
 \label{k31table}
\end{table}%
\begin{table}[t] 
\centering
\begin{tabular}{l l l}
&Energy levels for $k=5, M=2$&\\
\hline
Order &	$E_0$ & $E_1$ \\
\hline
$ \re^{-8 E /3}$ &   ${\underline{4.8544}694530582}$ & \underline{6.950123}71466050570772\\
$ \re^{-12 E /3}$ & ${\underline{4.854453}3648860}$ & \underline{6.950123641}54401512072\\
$\re^{-20 E /3}$ & ${\underline{4.854453651}9209}$ & \underline{6.9501236417927}2664768 \\
$\re^{-28 E /3}$ & ${\underline{4.85445365153}25}$  &\underline{6.95012364179271448}721\\
\hline
Numerical value & 4.8544536515315 &   6.95012364179271448613
\end{tabular}
\caption{The first two energy levels for $k=5, M=2$ calculated analytically from (\ref{exact-qc}). 
}
 \label{k52table}
\end{table}%

Let us now give some concrete results for the correction series (\ref{le-series}). In the case of ABJM theory, with $M=0$, 
the first few corrections read
\be\ba 
\lambda_1&=\frac{1}{\pi }\sin \left(16 x \right) ,\\
\lambda_2&=-\frac{4}{\pi} \sin ^2(4 x) \sin (20 x) \csc ^2(2 x),\\
\lambda_3&= \frac{8}{\pi} \sin (24 x)\csc ^2(2 x) \left(3 \sin ^2(4 x) \sin ^2(6 x)+\sin ^2(2 x) \sin ^2(8 x)+\sin ^2(10 x)\right)  ,\\
\lambda_4&= -\frac{8}{\pi}  \sin^2(4x)\sin (28x) \csc^2(2x) \bigl(
23+22\cos(4x)+19\cos(8x)+4\cos(12x)+3\cos(16x) \bigr).~~
\ea \ee
The results for $\lambda_1,\lambda_2$ and $\lambda_3$ are in perfect agreement with the results of \cite{huang} based on numerical fitting. 
We can check as well the higher order corrections by performing a detailed numerical analysis. First of all, we compute the first two energy levels 
in (\ref{inteq}), with high precision. To do this, we follow a procedure inspired by the analysis of \cite{hmo} and summarized in 
Appendix \ref{appendixnum}. On the other hand, we use (\ref{Volc}) to compute the corrections to the quantum volume up to 
\be \lambda_{10} \re^{-{52 \over k}E_{\rm eff}}.\ee 
The results are  shown in tables \ref{k03table} and \ref{k0table}, for $k=3,5,6$. As expected, the more instanton corrections we include 
in the analytic computation, the better we approach the numerical value. This can be seen in detail by considering the quantity 
\be \label{diff} \Delta_0^{(k)}(m)= \log_{10}\left| E^{\rm num}_0(k)-E^{(m)}_0(k) \right|,
\ee
where $E^{\rm num}_0$ is the numerical value of the ground state energy, and $E^{(m)}_0$ is the value computed from (\ref{Volc}) 
by including the first $m$ instanton corrections. As shown in \figref{Figos}, $E^{(m)}_0$ converges to $E^{\rm num}_0$ as 
$m$ grows. However, it does not converge monotonically, in contrast to what happened in the numerical analysis of \cite{km} for $k=1,2$ and $M=0$. 
In tables \ref{k31table}, \ref{k52table} we give additional numerical evidence for the validity of the quantization condition in the ABJ case with $M \neq 0$. 

\sectiono{A case study with $\mathcal{N}=6$ supersymmetry}

As emphasized in \cite{ghm}, the quantization condition studied in the previous section is a consequence of a stronger result, namely our 
explicit formula (\ref{spec-det}) for the spectral determinant. In principle, using this formula, one can compute the canonical 
partition functions by performing an expansion around $\kappa=0$ as in (\ref{gen-fun}). 
In \cite{cgm} this was checked in detail in the maximally supersymmetric cases, by using the computations of the partition functions in 
\cite{py,hmo, hmo2, honda-o,abj-moriyama}. In the case with maximal supersymmetry, the generalized theta function becomes a standard Jacobi 
theta function, and the higher genus contribution to the modified grand potential vanish, so the analysis of the spectral determinant 
is relatively straightforward. 

In this section we analyze in detail a case with $\CN=6$ supersymmetry, namely ABJM theory with $k=4$. This case is slightly richer than the maximally supersymmetric cases because the grand potential involves the all genus worldsheet instantons. However, the generalized theta function of this theory becomes a conventional theta function, as in the 
maximally supersymmetric cases. Therefore this case is not the most generic one, but it is a good starting point to start exploring the spectral determinants of theories with $\CN=6$ supersymmetry.


It follows from (\ref{gpmueff}) that the grand potential of ABJM theory with $k=4$ is given by 
\be \ba \label{J4}J(\mu, 4)= J^{({\rm p})}(\mu_{\rm eff},4)+J^{\rm WS}(\mu_{\rm eff},4)+ \mu_{\rm eff}
\widetilde{J}_b(\mu_{\rm eff},4)+\widetilde{J}_c(\mu_{\rm eff},4). \ea\ee
To calculate this quantity we have to take the limit $k\rightarrow 4$ in the general expression and be careful with the poles, 
as in \cite{cgm}. These however cancel, as we 
recalled above, so we can compute (\ref{J4}) by considering only its finite part. In particular 
$\widetilde{J}_b(\mu_{\rm eff},4)$ has no finite part, as shown in appendix \ref{annexj}, 
while the finite part of $\widetilde{J}_c(\mu_{\rm eff},4)$ is
\be  {1\over 12}\left(-\log \left( 1- 16 y^2\right)+\tilde{\varpi}_1(-y^2)\right), \quad y=\re^{-\mu},
\ee
where
\be 
\tilde{\varpi}_1(z)= -4 z \, _4F_3\left(1,1,\frac{3}{2},\frac{3}{2};2,2,2;-16 z\right).\ee
Let us now look at the worldsheet instanton part. For general $k$ and $M=0$  the expression (\ref{JWS}) reads
\be J^{\rm WS}(\mu, k)= \sum_{g \geq 0}\sum_{w,d\geq 1}{1\over w}(-1)^{dw} n_g^d \left( 2 \sin {2 \pi w\over k}\right)^{2g-2}\re^{-4 dw \mu/k}. \ee
It is convenient to split the sum over $w$ into even and odd part. This leads to
\be  \ba J^{\rm WS}(\mu, k)&=  \sum_{g \geq 0}\sum_{ w,d\geq 1}{1\over 2 w} n_g^d \left( 2 \sin {2 \pi w\over (k/2)}\right)^{2g-2}\re^{-4 dw \mu/(k/2)}\\
&\quad+\sum_{g \geq 0}\sum_{ w,d\geq 1}{1\over 2 w-1}(-1)^d n_g^d \left( 2 \sin {2 \pi (2w-1)\over k}\right)^{2g-2}\re^{-4 d(2w-1) \mu/k} .\ea \ee 
When $k=4$, the first part is (up to a factor $1/2$) precisely the worldsheet instanton part of the modified grand potential for the maximally supersymmetric theory 
$(k,M)=(2,1)$ analyzed in \cite{cgm}. The second part is a non trivial quantity which contains all genus contributions. We can then write the worldsheet instanton part of (\ref{J4}) as
\be J^{\rm WS}(\mu_{\rm eff}, 4)={1\over 2} J^{\rm WS}(\mu_{\rm eff}, 2,1)+I(\mu_{\rm eff}). \ee
The first term  was computed in \cite{cgm} and reads
 \be \label{jnf2}
J^{\rm WS}(\mu, 2,1)=   
{\mu_{\rm eff}^2 \over 2 \pi^2} \partial^2_t F^{\rm inst}_0(t ) - {\mu_{\rm eff} \over 2 \pi^2} \partial_t  F^{\rm inst}_0(t ) +{1\over 4 \pi^2} F^{\rm inst}_0(t) + F_1^{\rm inst}(t ).
\ee
Here we used $t= 2 \mu_{\rm eff}$ and 
\be  F^{\rm inst}_0(t ) =\sum_{d,w}n_0^d{1\over w^3} \re^{-  dw t }, \quad  F^{\rm inst}_1(t) =\sum_{d,w}\left({ n_0^d \over 12 }+n_1^0\right) \re^{-  dw t} . \ee
The second part of  $J^{\rm WS}$ contains contributions at all genera and can be written as
\be I(\mu_{\rm eff})=\sum_{g \geq 0}\sum_{ w,d\geq 1}{(-1)^d n_g^d \over 2 w-1} 4^{g-1}\re^{-d(2w-1) \mu_{\rm eff}} =\sum_{g\geq 0} \sum_{d \geq 1}4^{g-1} (-1)^d n^d_g \tanh ^{-1}\left(\re^{-d \mu_{\rm eff} }\right).\ee
By looking at the small $y$ expansion of this quantity we conjecture that it can be resummed in closed form in terms of an elliptic integral of the first kind, 
\be \label{allgenus}I(\mu_{\rm eff})=-\frac{1}{4} \log \left(\frac{2 K\left(16 y^2\right)}{\pi }\right)+{1\over 4 }\tanh ^{-1}\left(4y\right).\ee
We have checked the above equality order by order in a series expansion at small $y$, but we do not have a general proof of it. 
However, we will see that this conjecture reproduces the correct $Z(N,4)$ appearing in the large $y$ expansion of the spectral determinant (\ref{ZJ}).
This strongly suggests that (\ref{allgenus}) is a true identity. The existence of an identity like this is remarkable, since it resums the all-genus Gopakumar--Vafa expansion of the 
free energy. This resummation is needed if we want to reproduce the canonical partition functions: this requires an expansion around $y \rightarrow \infty$, while the original Gopakumar--Vafa 
expansion only holds at large radius, i.e. for $y\rightarrow 0$. 

By using (\ref{allgenus}) one finds 
\be \ba \label{j4lr}J(\mu, 4)&= A(4,0)+F_1(t) +F_1^{\rm NS}(t)+{\mu \over 4}+\frac{1}{4} \tanh ^{-1}\left(4\re^{-\mu}\right)
\\ &\quad+{1\over 8 \pi^2}\left(F_0(t)-t\partial_t F_0(t)+{1\over 2}t^2\partial_t^2 F_0(t)\right), \ea\ee
where 
\be \ba  F_0(t)&={t^3\over 6}+F_0^{\rm inst}(t), \\
 F_1(t)& = -{1\over 12} \log \left[ 64 y^2 (1-16 y^2) \right] -{1\over 2
} \log \left( {K(16y^2) \over \pi} \right),\\
F_1^{\rm NS}(t)&=-{1\over 24} \log{(1-16 y^2)\over y^4}. \ea\ee
The constant $A(4,0)$ is the standard constant map contribution of ABJM, 
whose exact value is given by
\be
A(4,0)=-\frac{\zeta (3)}{4 \pi ^2}-\frac{\log (2)}{2} .
\ee
The derivative of the grand potential takes the following closed form,
\be \label{DJ4}\partial_{\mu}J(\mu, k=4)=\frac{\mu_{\rm eff}^2}{8 \left(1-16 y^2\right) K\left(16 y^2\right)^2}-\frac{E\left(16 y^2\right)}{2 \left(16 y^2-1\right) K\left(16 y^2\right)}+\frac{1-2 y}{8 y-2}+{1 \over 4}.\ee
The large $\mu$ expansion of (\ref{j4lr}) reads
\be \ba J(\mu, 4)&=\frac{\mu ^3}{6 \pi ^2}+\frac{\mu }{4}-\frac{\zeta (3)}{4 \pi ^2}-\frac{\log (2)}{2} +y+\frac{\left(-4 \mu ^2-2 \mu -1\right) y^2}{2 \pi ^2}\\
&\quad+\frac{16 y^3}{3}+\frac{\left(-208 \mu ^2-4 \mu +32 \pi ^2-9\right) y^4}{16 \pi ^2}+\frac{256 y^5}{5}+ \mathcal{O}(y^6).\ea\ee
%

%
Once we know the modified grand potential, we can use (\ref{JXi}) and (\ref{JJ}) to obtain the grand canonical partition function 
or spectral determinant:
\be \label{Xi4} \Xi(\kappa,4) = \re^{J(\mu,4) } \vartheta_3 \left({1\over 2}\left(\xi-{\tau \over  4}-{1\over 12}\right),{\tau \over 2}\right), \ee
where
\be \tau={2 \ri \over \pi}\partial_t^2 F_0(t), \quad \xi={1\over 2 \pi}\left( t \partial_t^2 F_0(t)- \partial_t F_0(t)\right). \ee
Note that, although this theory is not maximally supersymmetric, the generalized theta function becomes in this case a Jacobi theta function. 
As explained in \cite{cgm, ghm} the spectrum is determined by the zeros of the theta function, and we find the exact quantization condition
\be  
\label{qc4}{\xi}-{\tau \over 4}-{1\over 12}=2n+1, \quad n=0,1,2, \cdots \ee
Interestingly, the quantum volume for $k=4$ is exactly related to that for $k=2$ and $M=1$ 
\be \label{Vol4} \text{Vol} (E,k=4)={1\over 2}\left({\xi}-{\tau \over 4}-{1\over 12}\right){16\pi^2}= \text{Vol}(E, k=2,M=1)+4 \pi ^2, \ee
where $\text{Vol}(E, k=2,M=1)$ is the quantum volume for the maximally supersymmetric theory with $k=2$, $M=1$ obtained in \cite{cgm}. 
Notice that, as we mentioned before, in this case the approximate quantization condition of \cite{km} is exact and does not need additional corrections. 

By following the arguments of \cite{cgm}, we can now use modular properties and analytic continuation to write the special determinant (\ref{Xi4}) in the orbifold frame, i.e. in the 
region $\mu \rightarrow -\infty$, where we make contact with the expansion (\ref{gen-fun}). This will allow us to compute the exact values of $Z(N,4)$ for finite $N$, 
which provides a check of the formula (\ref{spec-det}). In order to proceed we introduce the orbifold periods \cite{dmp}:
\be 
\label{standard}
\ba
\lambda&=\ri {\kappa \over 8 \pi}  {~}_3F_2\left(\frac{1}{2},\frac{1}{2},\frac{1}{2};1,\frac{3}{2};\frac{\kappa^2
   }{16}\right), 
 \\
 \partial_\lambda \CF_0 (\lambda)&= \ri { \kappa \over 4 } G^{2,3}_{3,3} \left( \begin{array}{ccc} {1\over 2}, & {1\over 2},& {1\over 2} \\ 0, & 0,&-{1\over 2} \end{array} \biggl| {\kappa^2\over 16}\right)- { \pi^2  \kappa\over 2}   {~}_3F_2\left(\frac{1}{2},\frac{1}{2},\frac{1}{2};1,\frac{3}{2};\frac{\kappa^2
   }{16}\right).   \ea
\ee
Here, $\CF_0(\lambda)$ is the genus zero free energy in the orbifold frame, normalized in such a way that is expansion around $\lambda=0$ is given by 
\be
\CF_0(\lambda)=-4 \pi ^2 \lambda ^2 \left(\log (2 \pi  \lambda )-\frac{3}{2}-\log (4)\right)+\cdots
\ee
Using modular properties of the periods and analytic  continuation  we find

\be\ba \label{Xi4o} \Xi(\kappa,4) &=\exp \biggl[\frac{1}{4} \tanh ^{-1}\left(\frac{\re^{\mu} }{4}\right)+{\mu \over 4}-{1\over 2 \pi^2}  \left( \CF_0(\lambda)- \lambda \partial_{\lambda}\CF_0(\lambda)+{1\over 2}\lambda^2  \partial^2_{\lambda}\CF_0(\lambda)\right) \biggr] \\
& \times \exp \biggl[\CF_1+F_1^{\rm NS}-{\ri \pi \over 8}+ \biggl] \vartheta_3 \left(\bar{\xi}-{1\over 4},2 {\bar{\tau}}\right),\ea\ee
where
\be
 \bar \tau=-{1\over 8 \pi^3 \ri}\partial_\lambda^2 \CF_0 (\lambda),\qquad 
 \bar \xi = {\ri \over 4 \pi^3} \left( \lambda \partial_\lambda^2 \CF_0 (\lambda) - \partial_\lambda \CF_0 (\lambda)\right).
\ee
The genus one free energy in the orbifold frame is given by
\be
\label{F1st} \CF_1=- \log \eta\left( 2 \bar \tau\right)-{1\over 2}\log 2. \ee
The small $\kappa$ expansion of (\ref{Xi4}) is now straightforward and one finds
\be 
\ba  \CF_1+F_1^{\rm NS}(t)-{\ri \pi \over 8}+\frac{1}{4} \tanh ^{-1}\left(\frac{\re^{\mu} }{4}\right)+{\mu \over 4}& =\frac{\kappa }{16}+\frac{\kappa ^3}{768}+\frac{\kappa ^4}{32768}+\mathcal{O}(\kappa^5),\\
 -{1\over 2 \pi^2}  \left( \CF_0(\lambda)- \lambda \partial_{\lambda}\CF_0(\lambda)+{1\over 2}\lambda^2  \partial^2_{\lambda}\CF_0(\lambda)\right) &=-\frac{\kappa ^2}{64 \pi ^2}-\frac{\kappa ^4}{3072 \pi ^2}+\mathcal{O}(\kappa^5),\\
\vartheta_3 \left(\bar{\xi}-{1\over 4},2 {\bar{\tau}}\right)&=1-\frac{\kappa ^3}{256 \pi }+\mathcal{O}(\kappa^5).\ea
\ee
Hence we have
\be 
 \Xi(\kappa, 4)= 1+ {\kappa \over 16}+\frac{\pi ^2-8}{512 \pi ^2}\kappa^2 + \mathcal{O}(\kappa^3),
 \ee
which reproduces the computation of the very first $Z(N,4)$ in \cite{hmo2}. Of course we can push the computation at higher order in $\kappa$ 
and reproduce all known $Z(N,4)$ for higher $N$.
  
\sectiono{Factorization of the spectral determinant}

 Since the potential $V_M(x)$ appearing in the ABJ(M) spectral problem is even, the eigenfunctions $\phi_n(x)$ (\ref{inteq}) can be classified according to their parity, as in ordinary 
 one-dimensional quantum-mechanical problems. The even energy 
 levels correspond to even eigenfunctions, while the odd energy levels correspond to odd eigenfunctions. Therefore, we can 
 split the spectral determinant (\ref{spectdet}) according to the 
 parity of the eigenfunctions, and we define
\be\label{Xeo}
\ba
\Xi_+(\kappa,k,M)=\prod_{n=0}^\infty (1+\kappa \re^{-E_{2n}}) ,\qquad
\Xi_-(\kappa,k,M)=\prod_{n=0}^\infty (1+\kappa \re^{-E_{2n+1}}).
\ea
\ee
If we introduce the operators with even/odd parity 
\be
\rho_\pm(x_1,x_2)=\frac{\rho(x_1,x_2) \pm \rho(x_1,-x_2)}{2},
\ee
the spectral determinants can be also written as
\be
\Xi_\pm(\kappa,k,M)=\det(1+\kappa \hat \rho_\pm)=\exp \left[ -\sum_{n=1}^\infty \frac{(-\kappa)^n}{n} \Tr \hat  \rho_\pm^n \right].
\ee
Notice that by construction, one immediately gets
\be
\Xi(\kappa,k,M)=\Xi_+(\kappa,k,M)\Xi_-(\kappa,k,M).
\ee
In this section we present an exact expression for the spectral determinants (\ref{Xeo}) in the case of $k=1, M=0$. We do not have a 
first principles derivation of such expressions, so we postulate a form that can be subsequently verified in detail. The expressions we propose are the following, 
\be\label{X1exact}
\ba
\Xi_+(\kappa,1)&=e^{J_+(\mu,1)}\vartheta_3\( \frac{1}{2}\(\frac{\xi}{2}+\frac{5}{24} \),\frac{\tau}{2} \) ,\\
\Xi_-(\kappa,1)&=e^{J_-(\mu,1)}\vartheta_4\( \frac{1}{2}\(\frac{\xi}{2}+\frac{5}{24} \),\frac{\tau}{2} \). 
\ea
\ee
In these expressions, 
\be
\ba
J_+(\mu,1)&=\frac{f_0(\mu)}{2}+f_1(\mu)+\frac{7}{16}\mu+A_+(1)
+\frac{1}{8} \log \( \frac{1+2\sqrt{2}e^{-\mu}+4e^{-2\mu}}{1-2\sqrt{2}e^{-\mu}+4e^{-2\mu}} \), \\
J_-(\mu,1)&=\frac{f_0(\mu)}{2}+f_1(\mu)-\frac{1}{16}\mu+A_-(1)
-\frac{1}{8} \log \( \frac{1+2\sqrt{2}e^{-\mu}+4e^{-2\mu}}{1-2\sqrt{2}e^{-\mu}+4e^{-2\mu}} \),
\ea
\ee
where
\be\label{f0f1}
\ba
f_0(\mu)&= \frac{1}{16\pi^2}\(F_0(t)-t \pd_t F_0(t) +\frac{t^2}{2}\pd_t^2 F_0(t)\) ,\\ 
f_1(\mu)&= F_1(t)+F_1^{\rm NS}(t)
\ea
\ee
and 
\be
A_+(1)=-\frac{\zeta(3)}{16\pi^2}-\frac{3}{8}\log2,\qquad
A_-(1)=-\frac{\zeta(3)}{16\pi^2}+\frac{5}{8}\log2,
\ee
In the $k=1$ context we should use $ t=4\mu_{\rm eff} $ and 
\be  
F_0(t ) ={t^3 \over 6}+\sum_{d,w}n_0^d{(-1)^{dw}\over w^3} \re^{-  dw t } . \ee
The standard and refined genus one free energies are given in terms of $z=\re^{-4\mu}$ as
\be\label{genus1}\ba 
F_1(t)= & -{1\over 12} \log \left[ 64 z (1+16 z) \right] -{1\over 2
} \log \left( {K(-16z) \over \pi} \right), \\
F^{\rm NS}_1 (t)= & {1\over 12} \log z -{1\over 24} \log(1+16 z). 
\ea
\ee
%
%
%
As a first check of the proposal \eqref{X1exact}, let us show that the above expressions lead to the right quantization conditions, i.e. $\Xi_\pm (\kappa,1)$ vanish when $\mu=E_{2n}+\ri \pi$, $\mu=E_{2n+1}+\ri \pi$, 
respectively. Let us first recall the quantization condition for $k=1$ \cite{cgm} 
\be\label{k1lev}
 {\xi \over 2}-{1\over 24} =m+{3 \over 4}, \quad m \geq 0.
 \ee
It is easy to see that the zeros of  $\Xi_{\pm}(E+\ri \pi,1)$ are given by
\be \ba {\xi \over 2}-{1\over 24}= & 2 m+{3 \over 4}, \quad m \geq 0,\\
 {\xi \over 2}-{1\over 24}=&2(m+1)+{3 \over 4}, \quad m \geq 0,
\ea\ee  
which leads precisely  to the odd and even energy levels for $k=1$ determined by (\ref{k1lev}). 

As a second check, one can verify that
\be
\Xi(\kappa,k)=\Xi_+(\kappa,k)\Xi_-(\kappa,k).
\ee
It is important to notice  that the total grand potential $J(\mu,1)$ differs from the sum
\be J_+(\mu,1)+J_-(\mu,1) \ee
by a term involving  the genus one free energy. More precisely one has
\be
J(\mu,1)=J_+(\mu,1)+J_-(\mu,1)-F_1(t)-F_1^{\rm NS}(t).
\ee
This additional contribution  comes from the product of the two theta functions
in $\Xi_\pm(\kappa,1)$.%
\footnote{We have used the identity
\[\vartheta_3 \(\frac{v}{2}, \frac{\tau}{2}\)\vartheta_4 \(\frac{v}{2}, \frac{\tau}{2}\)
=\vartheta_4(0,\tau)\vartheta_4(v,\tau).\]
The factor $\vartheta_4(0,\tau)$ contributes to the modified grand potential $J(\mu,1)$.
}

The third test concerns the large $\mu$ expansion for $J_{\pm}(\mu,1)$. If we write
\be
\ba
J_+(\mu,1)&=\frac{\mu^3}{3\pi^2}+\frac{7}{16}\mu+A_+(1)+J_+^\text{np}(\mu,1) ,\\
J_-(\mu,1)&=\frac{\mu^3}{3\pi^2}-\frac{1}{16}\mu+A_-(1)+J_-^\text{np}(\mu,1),
\ea
\ee
we find from the above expressions, 
\be
\ba
J_+^\text{np}(\mu,1)&=\frac{1}{\sqrt{2}}\re^{-\mu}-\frac{4}{3\sqrt{2}}\re^{-3\mu}
+\frac{16\mu^2+4\mu+1}{8\pi^2}\re^{-4\mu}-\frac{16}{5\sqrt{2}}\re^{-5\mu}
+\frac{64}{7\sqrt{2}}\re^{-7\mu}\\
&\quad+\biggl[ -\frac{52\mu^2+\mu/2+9/16}{4\pi^2}+2 \biggr] \re^{-8\mu}
+\frac{256}{9\sqrt{2}} \re^{-9\mu}+\cO(\re^{-11\mu}),\\
J_-^\text{np}(\mu,1)&=-\frac{1}{\sqrt{2}}\re^{-\mu}+\frac{4}{3\sqrt{2}}\re^{-3\mu}
+\frac{16\mu^2+4\mu+1}{8\pi^2}\re^{-4\mu}+\frac{16}{5\sqrt{2}}\re^{-5\mu}
-\frac{64}{7\sqrt{2}}\re^{-7\mu}\\
&\quad+\biggl[ -\frac{52\mu^2+\mu/2+9/16}{4\pi^2}+2 \biggr] \re^{-8\mu}
-\frac{256}{9\sqrt{2}} \re^{-9\mu}+\cO(\re^{-11\mu}).
\ea
\ee
These expansions can be reproduced from the expressions for the spectral traces
of $\rho_\pm$, which were found in \cite{hmo} up to $n=8$. The very first few values are given by
\be\label{Traces}
\ba
\Tr \rho_+&=\frac{1}{4 \sqrt{2}}, \\
\Tr \rho_+^2&=\frac{1}{16 \pi },\\
\Tr \rho_+^3&=\frac{3-2 \sqrt{2}}{64} ,\\
\Tr \rho_+^4&=\frac{1}{512} \left(1-\frac{8}{\pi ^2}\right),
\ea
\qquad\quad
\ba
\Tr \rho_-&=\frac{1}{4}-\frac{1}{4\sqrt{2}}, \\
\Tr \rho_-^2&=\frac{-3+\pi }{16 \pi },\\
\Tr \rho_-^3&=\frac{-12+\pi +2 \sqrt{2} \pi }{64 \pi }, \\
\Tr \rho_-^4&=-\frac{8+32 \pi -11 \pi ^2}{512 \pi ^2}.
\ea
\ee 
In fact, we can relate our factorized spectral determinants $\Xi_\pm$ to these spectral traces directly, by expressing them in terms of  orbifold quantities, like in \cite{cgm}. 
As shown in appendix \ref{appendixsd} in detail, one finds the following small $\kappa$ expansion:
\be
\log \Xi_\pm (\kappa,1)=-\sum_{n=1}^\infty \frac{(-\kappa)^n}{n} \Tr \rho_\pm^n,
\ee
which reproduces the exact values of $\Tr \rho_\pm^n$.
We have indeed computed $\Tr \rho_\pm^n$ up to $n=44$ and compared them
with the ones obtained from the orbifold expansion. The results show a perfect agreement.
 
\sectiono{Exact functional equations}
In the previous section, we have considered the factorization of the spectral determinant.
The reason why we focus on such a factorization is because the factorized spectral determinants $\Xi_\pm(\kappa,k,M)$
have a very rich structure.
In particular, these functions satisfy a set of exact functional equations as we will see in this section.
A similar property has already been found in the context of the spectral problem of certain ordinary differential equations
(see \cite{voros, dt, blz5} for example), where it indicates an unexpected connection to integrable models.
Our result extends this type of properties to the spectral problem of the Fredholm integral equation \eqref{inteq}.
We hope that our findings may give a clue of a connection to some integrable systems.

\subsection{Wronskian-like relations}

We consider the spectral determinant \eqref{spectdet} and the factorized ones \eqref{Xeo}.
%
Remarkably, these functions satisfy the following beautiful equations for given $k$ and $M$:
\be \label{FE1}
\ba
&\re^{\frac{M}{2k}\pi \ri} \Xi_+\(\ri \kappa,k,M+1\)\Xi_-\(-\ri \kappa,k, M-1\)\\
&\quad -\re^{-\frac{M}{2k}\pi \ri} \Xi_+\(-\ri \kappa,k, M+1\)\Xi_-\(\ri \kappa,k, M-1\)
=2\ri \sin\( \frac{M\pi}{2k} \) \Xi\(\kappa,k, M\) ,
\ea
\ee
and
\be \label{FE2}
\ba
&\re^{-\frac{M}{2k}\pi \ri} \Xi_+\(\ri \kappa,k,M-1\)\Xi_-\(-\ri \kappa,k, M+1\)\\
&\quad +\re^{\frac{M}{2k}\pi \ri} \Xi_+\(-\ri \kappa,k, M-1\)\Xi_-\(\ri \kappa,k, M+1\)
=2\cos\( \frac{M\pi}{2k} \) \Xi\(\kappa,k, M\) .
\ea
\ee
The equation \eqref{FE1} is quite similar to the so-called quantum Wronskian relation \cite{blz2}.
We do not have a general proof for these relations but we can test them by computing the spectrum and its spectral traces 
from the quantization condition (\ref{exact-qc}), and by doing small $\kappa$ expansions of the spectral determinants. This can be done for various values of the coupling $k$.
Such tests strongly suggest that the functional equations \eqref{FE1} and \eqref{FE2} are widely valid not only for integral values of $k$, but also for non-integral values.

In particular, if $k$ is an integer, the equations \eqref{FE1} and \eqref{FE2} are essentially equivalent due to the Seiberg-like duality of ABJ theories \cite{abj}.
This duality relates the partition function for $(k,M)$ to the one for $(k,k-M)$.
In terms of the spectral determinants, it simply says that
\be
\Xi(\kappa,k,M)=\Xi(\kappa,k,k-M),\qquad \Xi_\pm(\kappa,k,M)=\Xi_\pm(\kappa,k,k-M).
\ee
If one considers the case $M=k-m$ in \eqref{FE2}, one gets
\be \label{FE3}
\ba
&-\ri \re^{\frac{m}{2k}\pi \ri} \Xi_+\(\ri \kappa,k,k-m-1\)\Xi_-\(-\ri \kappa,k, k-m+1\)\\
&\quad +\ri \re^{-\frac{m}{2k}\pi \ri} \Xi_+\(-\ri \kappa,k, k-m-1\)\Xi_-\(\ri \kappa,k, k-m+1\)
=2\sin\( \frac{m\pi}{2k} \) \Xi\(\kappa,k, k-m\) .
\ea
\ee
Using the Seiberg-like duality $\Xi_\pm(\kappa,k,k-m\pm 1)=\Xi_\pm(\kappa,k,m \mp 1)$,
it is easy to see that this equation is equivalent to \eqref{FE1} for $M=m$.
We note that for non-integral $k$, the equations \eqref{FE1} and \eqref{FE2} are independent.

Moreover we  conjecture another functional equation, which associates $\Xi_\pm(\kappa,k,1)$ to $\Xi(\kappa,k)$. 
\be \label{id1}
\Xi_+(\ri \kappa, k, 1)\Xi_+(-\ri \kappa, k, 1)+\frac{\kappa}{4k} \Xi_-(\ri \kappa, k, 1)\Xi_-(-\ri\kappa,k, 1)= \Xi(\kappa,k), \quad \forall k. 
\ee
As before the identity (\ref{id1}) can be checked by  computing the spectrum and by doing a small  $\kappa$ expansion.
We tested this relation for various values of the coupling $k$,
and conjecture that it is valid for any $k$.

Let us comment on a consequence of these functional equations.
For odd $k$, there are $k+1$ independent functions $\Xi_\pm(\kappa,k,M)$, ($M=0,\dots, \frac{k-1}{2}$)
due to the Seiberg-like duality. 
A simple counting shows that the functional equation \eqref{FE1} (or equivalently \eqref{FE2}) gives $k-1$ independent
constraints.
This means that if we know the two functions $\Xi_\pm (\kappa,k)$ in the ABJM theory,
all the other functions $\Xi_\pm (\kappa,k, M)$ in ABJ theory are determined by the functional equations.
Similarly, for even $k$, the functional equation \eqref{FE1} gives $k-1$ independent constraints among
the $k+2$ independent functions $\Xi_\pm(\kappa,k,M)$ ($M=0,\dots,\frac{k}{2}$).
In this case, the equation \eqref{FE1} only does not determine the ABJ spectral determinants from the ABJM ones.
Since we have the additional relation \eqref{id1}, one might expect that combining these equations, the ABJ spectral determinants
are fixed, as for odd $k$.
However, this is not the case. One can check that the equations \eqref{FE1} and \eqref{id1} are not sufficient to determine
all the ABJ spectral determinants only from the ABJM ones.
We need more information for even $k$.\footnote{
If we additionally give the traces of the odd powers of $\rho_+$ for $M=1$, for example,
then the other traces in ABJ theories are fixed.
}

In the rest of this subsection we exploit  our exact solution (\ref{X1exact}) to further test \eqref{id1} in the case of $k=1$. 
More precisely, we are interested in studying the following identity:
\be \label{id1k1}
\Xi_+(\ri \kappa, 1)\Xi_+(-\ri \kappa, 1)+\frac{\kappa}{4} \Xi_-(\ri \kappa, 1)\Xi_-(-\ri \kappa,1)= \Xi(\kappa,1).
\ee
Notice that under $\kappa \to \pm \ri \kappa$, the chemical potential changes according to
\be
\mu \to \mu \pm \frac{\pi \ri}{2}.
\ee
Starting from (\ref{f0f1}) it is easy to see that
\be
\ba
f_0\( \mu + \frac{\pi \ri}{2} \)+ f_0\( \mu - \frac{\pi \ri}{2} \) &= 2f_0(\mu)-\frac{1}{4}\pd_t^2 {F}_0(t), \\
f_1\( \mu + \frac{\pi \ri}{2} \) + f_1\( \mu - \frac{\pi \ri}{2} \)  &= 2f_1(\mu).
\ea
\ee
It follows that
\be
\ba
J_+\( \mu + \frac{\pi \ri}{2} \) + J_+\( \mu - \frac{\pi \ri}{2} \) &= J(\mu,1)+F_1(t)+F_1^{\rm NS}(t)-\log 2+\frac{\mu}{2}-\frac{1}{8}\pd_t^2 {F}_0(t),\\
J_-\( \mu + \frac{\pi \ri}{2} \)+ J_-\( \mu - \frac{\pi \ri}{2} \) &= J(\mu,1)+F_1(t)+F_1^{\rm NS}(t)+\log 2-\frac{\mu}{2}-\frac{1}{8}\pd_t^2 {F}_0(t).
\ea
\ee
Similarly one has 
\be
\xi\( E \pm \frac{\pi \ri}{2} \)= \xi(E) \pm \frac{\tau}{2}-1,\qquad
\tau\( E \pm \frac{\pi \ri}{2} \)  = \tau(E) \mp 4.
\ee
Therefore,
\be
\ba
\vartheta_3\( \frac{1}{2}\( \frac{\xi\( E \pm \frac{\pi \ri}{2} \)}{2}+\frac{5}{24}\), \frac{\tau\( E \pm \frac{\pi \ri}{2} \)}{2} \)
&=\vartheta_3 \( \frac{\xi}{4}-\frac{7}{48} \pm \frac{\tau}{8}, \frac{\tau}{2} \),\\
\vartheta_3\( \frac{1}{2}\( \frac{\xi\( E \pm \frac{\pi \ri}{2} \)}{2}-\frac{19}{24}\), \frac{\tau\( E \pm \frac{\pi \ri}{2} \)}{2} \)
&=\vartheta_4 \( \frac{\xi}{4}-\frac{7}{48} \pm \frac{\tau}{8}, \frac{\tau}{2} \).
\ea
\ee
Using the identities
\be
\ba
\vartheta_3(x+y,\tau) \vartheta_3(x-y, \tau)&=\vartheta_3(2x,2\tau)\vartheta_3(2y,2\tau)+\vartheta_2(2x,2\tau)\vartheta_2(2y,2\tau),\\
\vartheta_4(x+y,\tau) \vartheta_4(x-y, \tau)&=\vartheta_3(2x,2\tau)\vartheta_3(2y,2\tau)-\vartheta_2(2x,2\tau)\vartheta_2(2y,2\tau),
\ea
\ee
we get
\be
\ba
&\vartheta_3 \( \frac{\xi}{4}-\frac{7}{48} + \frac{\tau}{8}, \frac{\tau}{2} \)
\vartheta_3 \( \frac{\xi}{4}-\frac{7}{48} - \frac{\tau}{8}, \frac{\tau}{2} \)=\\
&\vartheta_3\(\frac{\xi}{2}-\frac{7}{24},\tau\)\vartheta_3\( \frac{\tau}{4},\tau \)+\vartheta_2\(\frac{\xi}{2}-\frac{7}{24},\tau\)\vartheta_2\( \frac{\tau}{4},\tau \).\ea
\ee
Similarly 
\be
\ba
&\vartheta_4 \( \frac{\xi}{4}-\frac{7}{48} + \frac{\tau}{8}, \frac{\tau}{2} \)
\vartheta_4 \( \frac{\xi}{4}-\frac{7}{48} - \frac{\tau}{8}, \frac{\tau}{2} \)=\\
& \vartheta_3\(\frac{\xi}{2}-\frac{7}{24},\tau\)\vartheta_3\( \frac{\tau}{4},\tau \)-\vartheta_2\(\frac{\xi}{2}-\frac{7}{24},\tau\)\vartheta_2\( \frac{\tau}{4},\tau \).
\ea
\ee
It follows that
\be
\ba
\Xi_+(\ri \kappa,1)\Xi_+(-\ri \kappa,1)&=\frac{1}{2} \Xi(\kappa,1)\left(\vartheta_3\( \frac{\tau}{4},\tau \)+\frac{\vartheta_2(v,\tau)}{\vartheta_3(v,\tau)} \vartheta_2 \(\frac{\tau}{4},\tau\) \right)\\
& \times  \exp \biggl[F_1+F_1^\text{NS}+\frac{\mu}{2}-\frac{1}{8} \pd_t^2 F_0 \biggr]
,\\
\Xi_-(\ri \kappa,1)\Xi_-(-\ri \kappa,1)&=2 \Xi(\kappa,1)
\left(\vartheta_3\( \frac{\tau}{4},\tau \)-\frac{\vartheta_2(v,\tau)}{\vartheta_3(v,\tau)} \vartheta_2 \(\frac{\tau}{4},\tau\) \right)\\
& \times  \exp \biggl[F_1+F_1^\text{NS}-\frac{\mu}{2}-\frac{1}{8} \pd_t^2 F_0 \biggr].
\ea
\ee
By using the above expression one can write  (\ref{id1k1}) as
\be\label{cc}
F_1(t)+F_1^\text{NS}(t)+\frac{\mu}{2}-\frac{1}{8} \pd_t^2 F_0 (t)= -\log \vartheta_3\( \frac{\tau}{4},\tau \).
\ee
We have checked this identity order by order in the large $\kappa$ expansion, as well as numerically.
It would be interesting to confirm the functional equations \eqref{FE1} and \eqref{id1} at $k=2$ in a similar way
by using the exact solutions.

\subsection{Relations among different levels}
In addition to the general relations found in the previous subsection,
there are some accidental relations among the spectral determinants for different values of $k$.
We find that the following relations hold: 
\be
\ba
\Xi(\kappa,4)&=\Xi_+(\kappa,2,1) ,\\
\Xi(\kappa,4,1)&=\Xi_-(\kappa,2)=\Xi(\ri \sqrt{\kappa},1)\Xi(-\ri \sqrt{\kappa},1) ,\\
\Xi(\kappa,4,2)&= \Xi_-(\kappa,2,1), \\
\Xi(\kappa,8,2)&=\Xi_-(\kappa,4,2).
\ea
\label{Xi-relation}
\ee
These relations can be checked as follows.
Let us recall the relation \eqref{Vol4} for the quantum volumes.
Considering the quantization condition, the relation \eqref{Vol4} implies the equality
\be
E_n(k=4)=E_{2n}(k=2, M=1), \qquad n=0,1,2,\dots.
\ee
Thus we immediately find the first line in \eqref{Xi-relation} by definition.
Similarly, we find
\be
\ba
\Vol(E,4,1)&=\Vol(E,2)-4\pi^2=4\Vol\( \frac{E}{2},1 \), \\
\Vol(E,4,2)&=\Vol(E,2,1)-4\pi^2, \\
\Vol(E,8,2)&=\Vol(E,4,2)-8\pi^2,
\ea
\ee
From these relations, we find the relations on the energy levels, and then get \eqref{Xi-relation}.
As a further test, one can check the equalities around $\kappa=0$.
For example, the first and third lines in \eqref{Xi-relation} show that the spectral determinant for $k=2, M=1$
splits into two part
\be
\Xi(\kappa,2,1)=\Xi(\kappa,4)\Xi(\kappa,4,2).
\label{Xi21=Xi4Xi42}
\ee
One can check this equation by substituting the exact values of the partition function computed in \cite{hmo2, honda-o}.
Notice that we  already know the exact spectral determinants for $k=1,2,4$, $M=0$ as well as for $k=2$, $M=1$ \cite{cgm}.
These data fix $\Xi(\kappa,4,1)$ and $\Xi(\kappa,4,2)$ through \eqref{Xi-relation}.
For example, using \eqref{Xi21=Xi4Xi42}, we get
\be
\Xi(\kappa,4,2)=\frac{\Xi(\kappa,2,1)}{\Xi(\kappa,4)}
=\re^{J(\mu,2,1)-J(\mu,4)} \frac{\vartheta_3(\xi-\frac{\tau}{4}-\frac{7}{12},\tau)}
{\vartheta_3( \frac{1}{2}( \xi-\frac{\tau}{4}-\frac{1}{12}),\frac{\tau}{2}  ) } .
\ee
A simple calculation shows that this is written as
\be
\Xi(\kappa,4,2)=\re^{J(\mu,4,2)}\vartheta_4\( \frac{1}{2}\( \xi-\frac{\tau}{4}-\frac{1}{12}\),\frac{\tau}{2} \),
\ee
where
\be
J(\mu,4,2)=J(\mu,2,1)-J(\mu,4)+F_1(t)+F_1^\text{NS}(t), \qquad t=2\mu_{\rm eff}.
\ee
Similarly, $\Xi(\kappa,4,1)$ is fixed by the second line in \eqref{Xi-relation} by using the exact expression
for $\Xi(\kappa,1)$ in \cite{cgm}.

\section{Conclusion}
In this paper  we studied the spectral problem appearing in the Fermi gas formulation of ABJ(M) theory. By generalizing the recent study of maximally supersymmetric ABJ(M) theories in \cite{cgm}, 
we found an exact expression for the spectral determinant in terms of a generalized theta function, and then we computed the quantum volume by looking at the zeros of this spectral determinant. 
This method has the advantage of overcoming many technical difficulties encountered in \cite{km,kallen} 
and leads to an exact quantization condition for the spectrum. Our result also shows that the quantization conditions conjectured in \cite{km,kallen} are only approximate, 
although they become exact in the maximally supersymmetric cases. 
Our quantization condition agrees with a recent numerical analysis in \cite{huang}, and we tested it against a high precision, numerical computation of the spectrum. 
As an application of our results, we also conjectured some functional equations for the spectral determinants.
Note that the formalism we used in this paper is very powerful and completely general. As explained in \cite{ghm}, it also has important applications in topological string theory. 

This work can be extended in many ways. First of all, it would be interesting to understand the structure of the spectral determinant in other cases with $\CN=6$ supersymmetry. 
In the ABJM theory with $k=4$, we could resum the all-genus expansion of the modified grand potential in order to understand the expansion of the spectral determinant 
at small fugacity. It would be very interesting to understand if this resummation can be done in general. This will probably require a better understanding of the 
modular properties of the modified grand potential and of the generalized theta function at finite $k$. 

Another avenue to explore is the generalization of these results to other Chern--Simons--matter theories. This is not completely straightforward: 
although our results for the spectral determinant 
and the quantization conditions are quite general and can be easily extended to other models,  
our detailed computations rely on a detailed knowledge of the modified grand potential, which so far has been only achieved for ABJ(M) theory. 
Nevertheless, we hope that the results obtained in this paper will be useful to further understand the non-perturbative structure of Chern--Simons--matter theories and their large $N$ duals.

\section*{Acknowledgements}

We would like to thank Patrick Dorey, Masazumi Honda, Sanefumi Moriyama, Tomoki Nosaka, Kazumi Okuyama
and Roberto Tateo for useful discussions and correspondence.
The work of A.G. and M.M. is supported in part by the Fonds National Suisse, 
subsidies 200021-156995 and 200020-141329, and by the NCCR 51NF40-141869 ``The Mathematics of Physics" (SwissMAP). 

\appendix

\section{The finite part of $\tilde J_b(\mu_{\rm eff},k,M)$} \label{annexj}

We want to show that, when $k$ is integer, the coefficient $\widetilde{b}_\ell(k) $ defined in (\ref{blj}) has no finite part. 
More precisely, let us consider the expansion of $\widetilde{b}_\ell(k)$ around an integer $k_0$. Since $\widetilde{b}_\ell(k) $ has a simple pole there, one has 
\be\label{bljsp}
\widetilde{b}_\ell(k)={\tilde b_{\ell}^{-1}(k_0) \over k-k_0}+\tilde b_{\ell}^{0}(k_0) +\mathcal{O}(k-k_0). 
\ee
The finite part of $\widetilde{b}_\ell(k) $ as $k \rightarrow k_0$ is $\tilde b_{\ell}^{0}(k_0)$. We want to show that $\tilde b_{\ell}^{0}(k_0)=0$. 
Let us start by looking at the case in which $k_0$ is even. From (\ref{blj}) one can see that the finite part is
\be \label{b0k}\tilde b_{\ell}^{0}(k_0)= -\frac{\ell}{2\pi}\sum_{j_L,j_R}\sum_{\ell=dw}\sum_{d_1+d_2=d}N^{d_1,d_2}_{j_L,j_R}\frac{\ri m_L m_R(d_1-d_2) \cos \left(\frac{1}{2} \pi  k_0 w (d_1+d_2+m_L+m_R+1)\right)}{2 \pi  w^2}, \ee
where we have denoted
\be
m_{L,R}= 2 j_{L,R}+1. 
\ee
Since the BPS invariants of local $\IP^1 \times \IP^1$, $N^{d_1,d_2}_{j_L,j_R}$, are symmetric under the exchange $d_1\leftrightarrow d_2$, the above quantity vanishes. 
When $k_0$ is odd, it is convenient to split the sum over $w$ in (\ref{blj})  into  even  $w$ and odd  $w$. The sum over even $w$  can be reduced to (\ref{b0k}) and therefore vanishes.
The sum over odd $w$ gives instead a contribution of the form 
\be\label{woddb} 
 -\frac{\ell}{2\pi}\sum_{j_L,j_R}\sum_{\ell=dw}\sum_{d_1+d_2=d}N^{d_1,d_2}_{j_L,j_R} \frac{e^{\frac{1}{2} i \pi  (k (d_1 w-d_2 w-1)-w)} \sin \left(\frac{1}{2} \pi  k_0 m_L w\right) \sin \left(\frac{1}{2} \pi  k_0 w m_R\right)}{2 \pi  w^2}.\ee
For local $\mathbb{P}^1\times \mathbb{P}^1 $, the non-vanishing BPS invariants are such that 
\be m_L+m_R=2n+1, \quad n \in \mathbb{Z}.\ee
It follows that
\be \sin \left(\frac{1}{2} \pi  k_0 m_L w\right) \sin \left(\frac{1}{2} \pi  k_0 w m_R\right)=0.\ee
Hence, (\ref{woddb}) also vanishes. This proves our statement. 

\section{The spectral determinant at the orbifold point}\label{appendixsd}
Let us compute the expansions of the spectral determinants $\Xi_\pm(\kappa,1)$, given in (\ref{X1exact}), around the orbifold point $\kappa=0$. 
One finds
\be \ba  \Xi_+(\kappa,1)&=\exp\left[ \CF_1 +F_1^{\rm NS} - {1\over 8 \pi^2}  \left(  \CF_0(\lambda)- \lambda \partial_{\lambda}\CF_0(\lambda)+{\lambda^2\over 2}  \partial^2_{\lambda}\CF_0(\lambda)\right) \right] \\
  &  \times \exp\left[\frac{1}{8} \log \left(\frac{4+2 \sqrt{2}\kappa+\kappa^2}{4-2 \sqrt{2}\kappa+\kappa^2}\right) + \frac{7 \log (\kappa )}{16}-\frac{15}{8}  \log (2) \right]  \bar\theta^{(1)}(\xi,\tau),\\ 
\Xi_-(\kappa, 1)&= \exp\left[ \CF_1 +F_1^{\rm NS} - {1\over 8 \pi^2}  \left(  \CF_0(\lambda)- \lambda \partial_{\lambda}\CF_0(\lambda)+{\lambda^2\over 2}  \partial^2_{\lambda}\CF_0(\lambda)\right)  \right] \\
  &  \times \exp\left[ -\frac{1}{8}  \log \left(\frac{4+2 \sqrt{2}\kappa+\kappa^2}{4-2 \sqrt{2}\kappa+\kappa^2}\right) -\frac{\log (\kappa )}{16}-\frac{7}{8}  \log (2)\right]  \bar\theta^{(2)}(\xi,\tau),\ea\ee
where $F_1^{\rm NS}$ is given in (\ref{genus1}) and $\CF_1 , \CF_0$ can be obtained from (\ref{standard}), (\ref{F1st}) by replacing 
\be \kappa \rightarrow -\ri \kappa^2.\ee
In these equations, we defined
\be\ba \bar\theta^{(1)}(\xi,\tau)=&\left[\re^{\ri \pi \over 4} \vartheta_2\left({\xi \over 8 \tau}-{1\over 96 \tau}+{1\over 4}, 8 \tau\right)+\vartheta_2\left({\xi \over 8 \tau}-{1\over 96 \tau}, 8 \tau\right) \right. \\
& \left.- \ri \re^{\ri \pi \over 4} \vartheta_1\left({\xi \over 8 \tau}-{1\over 96 \tau}+{1\over 4}, 8 \tau\right)-\ri\vartheta_1\left({\xi \over 8 \tau}-{1\over 96 \tau}, 8 \tau\right) \right] ,\\
 \bar\theta^{(2)}(\xi,\tau)=&\left[\re^{-3\ri \pi \over 4} \vartheta_2\left({\xi \over 8 \tau}-{1\over 96 \tau}+{1\over 4}, 8 \tau\right)+\vartheta_2\left({\xi \over 8 \tau}-{1\over 96 \tau}, 8 \tau\right) \right. \\
& \left.- \ri \re^{-3\ri \pi \over 4} \vartheta_1\left({\xi \over 8 \tau}-{1\over 96 \tau}+{1\over 4}, 8 \tau\right)-\ri\vartheta_1\left({\xi \over 8 \tau}-{1\over 96 \tau}, 8 \tau\right) \right].\ea\ee
Let us look at the series expansion of the terms appearing in $\Xi_+(\kappa,1)$. We have:
\be
\ba
 \CF_1 +F_1^{\rm NS} +\frac{1}{8} \log \left(\frac{4+2 \sqrt{2}\kappa+\kappa^2}{4-2 \sqrt{2}\kappa+\kappa^2}\right)+\frac{ \log (\kappa )}{2} &=\frac{\kappa }{4 \sqrt{2}}-\frac{\kappa ^3}{48 \sqrt{2}}+ \mathcal{O}(\kappa^5),\\
  {1\over 8 \pi^2}  \left( - \CF_0(\lambda)+\lambda \partial_{\lambda}\CF_0(\lambda)-{\lambda^2\over 2}  \partial^2_{\lambda}\CF_0(\lambda)\right)&= \frac{\kappa ^4}{256 \pi ^2}-\frac{\kappa ^8}{12288 \pi ^2}+ \mathcal{O}(\kappa^{12}),\\
 \exp\left[ -\frac{15}{8}  \log (2)-\frac{\log (\kappa )}{16}\right]  \bar\theta^{(1)}(\xi,\tau)& =1-\frac{\kappa ^2}{32 \pi }  +\frac{\kappa ^3}{64}-\frac{\pi ^2-1} {2048 \pi ^2}\kappa ^4+ \mathcal{O}(\kappa^5).
\ea
\ee
This leads to
\be \log \Xi_+(\kappa,1)=\frac{\kappa }{4 \sqrt{2}}-\frac{\kappa ^2}{32 \pi }+\frac{3-2 \sqrt{2}}{192} \kappa ^3+\frac{1}{2048} \left(\frac{8}{\pi ^2}-1\right)\kappa ^4+ \mathcal{O}(\kappa^5).\ee
A similar computation holds for $\Xi_-(\kappa,1)$.

\section{Numerical calculation of the spectrum}\label{appendixnum}
In this Appendix we explain how to compute numerically the first two energy levels of the spectrum (\ref{inteq}), with high precision.
We introduce the following two functions:
\be
\ba
\phi_\ell^+(x)&=\frac{1}{\cosh (\frac{x}{2k})} \int_{-\infty}^\infty \frac{dx'}{2\pi k} 
\frac{\cosh (\frac{x'}{2k})}{2\cosh (\frac{x-x'}{2k})}  V_M(x') \phi_{\ell-1}^+(x'),\\
\phi_\ell^-(x)&=\frac{1}{\cosh (\frac{x}{2k})} \int_{-\infty}^\infty \frac{dx'}{2\pi k} 
\frac{\sinh (\frac{x'}{2k} ) \tanh (\frac{x'}{2k})}{2\cosh (\frac{x-x'}{2k})} V_M(x') \phi_{\ell-1}^-(x'),
\ea
\label{eq:int_eq}
\ee
where $V_M(x)$ is defined by \eqref{Vabj}.
As  shown in \cite{hmo, hmo2, honda-o}, these functions are building blocks to construct the matrix elements $\rho_\pm^\ell(x,y)$,
and one can compute the spectral traces $\Tr \rho_\pm^\ell$ from $\phi_\pm^\ell(x)$.
Recalling that the traces $\Tr \rho_\pm^\ell$ are also given by
\be
\Tr \rho_+^\ell=\sum_{n=0}^\infty \re^{-\ell E_{2n}},\qquad
\Tr \rho_-^\ell=\sum_{n=0}^\infty \re^{-\ell E_{2n+1}},
\ee
one can compute the first two energies levels from 
\be
\re^{-E_0}=\lim_{\ell \to \infty} \frac{\Tr \rho_+^{\ell}}{\Tr \rho_+^{\ell-1}}, \qquad
\re^{-E_1}=\lim_{\ell \to \infty} \frac{\Tr \rho_-^{\ell}}{\Tr \rho_-^{\ell-1}},
\ee
where we have used that
\be
\re^{-\ell E_0}>\re^{-\ell E_1}>\re^{-\ell E_2}>\dots .
\ee
From a practical point of view, there is another simpler way to compute $E_0$ and $E_1$.
This way is based on the observation that the two functions $\phi_\pm^\ell(0)$ already contain all the information of the spectrum.
Indeed we have
\be
\ba
\phi_\ell^+(0)&=C_0 \re^{-\ell E_0}+C_2 \re^{-\ell E_2}+C_4 \re^{-\ell E_4}+\dots, \\
\phi_\ell^-(0)&=C_1  \re^{-\ell E_1}+C_3\re^{-\ell E_3}+C_5 \re^{-\ell E_5}+\dots,
\ea
\label{eq:obs}
\ee
where $C_n$ are constant coefficients.
Using this observation, one immediately finds
\be\ba 
\re^{-E_0}=\lim_{\ell \to \infty} \frac{\phi_\ell^+(0)}{\phi_{\ell-1}^+(0)},\qquad
\re^{-E_1}=\lim_{\ell \to \infty} \frac{\phi_\ell^-(0)}{\phi_{\ell-1}^-(0)}.
\ea
\label{eq:E01}
\ee
These expressions are technically useful because we do not need to perform any integral over $x$.
The integral equations \eqref{eq:int_eq} can be solved numerically for given $k$ and $M$ with high precision.
Once we get $\phi_\pm^\ell(x)$ up to some values of $\ell$, we can estimate the energies by \eqref{eq:E01}.
This method is very powerful to compute $E_0$ and $E_1$ numerically.
In fact, we have checked that this method reproduces the energies computed from the exact quantization condition
in \cite{cgm} for $(k,M)=(1,0), (2,0), (2,1)$ with very high (at least 100-digit) precision.

\end{document}